\documentclass[aps,pra,twocolumn,superscriptaddress,showpacs,amsfonts]{revtex4-1}
\usepackage{epsfig}
\usepackage{graphicx}
\usepackage{amsmath}
\usepackage{amssymb}
\usepackage{color}
\usepackage{bm}

\newcommand{\beq}{\begin{equation}}
\newcommand{\eeq}{\end{equation}}
\newcommand{\beqarray}{\begin{eqnarray}}
\newcommand{\eeqarray}{\end{eqnarray}}

\allowdisplaybreaks

\begin{document}

\title{
Spin-Orbital Order Modified by Orbital Dilution in Transition Metal Oxides:\\
From Spin Defects to Frustrated Spins Polarizing Host Orbitals}

\author{Wojciech Brzezicki}
\affiliation{Marian Smoluchowski Institute of Physics, Jagiellonian University,
             prof. S. \L{}ojasiewicza 11, PL-30348 Krak\'ow, Poland }
\affiliation{CNR-SPIN, IT-84084 Fisciano (SA), Italy, and \\
             Dipartimento di Fisica \textquotedblleft{}E. R. Caianiello\textquotedblright{},
             Universit\'a di Salerno, IT-84084 Fisciano (SA), Italy}

\author{Andrzej M. Ole\'s}
\affiliation{Max-Planck-Institut f\"ur Festk\"orperforschung,
             Heisenbergstrasse 1, D-70569 Stuttgart, Germany}
\affiliation{Marian Smoluchowski Institute of Physics, Jagiellonian University,
             prof. S. \L{}ojasiewicza 11, PL-30348 Krak\'ow, Poland }

\author{Mario Cuoco}
\affiliation{CNR-SPIN, IT-84084 Fisciano (SA), Italy, and \\
             Dipartimento di Fisica \textquotedblleft{}E. R. Caianiello\textquotedblright{},
             Universit\'a di Salerno, IT-84084 Fisciano (SA), Italy}

\date{24 December 2014}

\begin{abstract}
We investigate the changes in spin and orbital patterns induced by 
magnetic transition metal ions without an orbital degree of freedom 
doped in a strongly correlated insulator with spin-orbital order. In 
this context we study the $3d$ ion substitution in $4d$ transition 
metal oxides in the case of $3d^3$ doping at either $3d^2$ or $4d^4$ 
sites which realizes \textit{orbital dilution} in a Mott insulator. 
Although we concentrate on this doping case as it is known 
experimentally and more challenging than other oxides due to finite 
spin-orbit coupling, the conclusions are more general. We derive the 
effective $3d-4d$ (or $3d-3d$) superexchange in a Mott insulator with 
different ionic valencies, underlining the emerging structure of the 
spin-orbital coupling between the impurity and the host sites and 
demonstrate that it is qualitatively different from that encountered in 
the host itself. This derivation shows that the interaction between the 
host and the impurity depends in a crucial way on the type of doubly 
occupied $t_{2g}$ orbital. One finds that in some cases, due to the 
quench of the orbital degree of freedom at the $3d$ impurity, the spin 
and orbital order within the host is drastically modified by doping. 
The impurity acts either as a spin defect
accompanied by an orbital vacancy in the spin-orbital structure when
the host-impurity coupling is weak, or it favors doubly occupied active
orbitals (orbital polarons) along the $3d-4d$ bond leading to
antiferromagnetic or ferromagnetic spin coupling. This competition
between different magnetic couplings leads to quite different ground
states. In particular, for the case of a finite and periodic $3d$ atom
substitution, it leads to striped patterns either with alternating
ferromagnetic/antiferromagnetic domains or with islands of saturated
ferromagnetic order. We find that magnetic frustration and spin 
degeneracy can be lifted by the quantum orbital flips of the host but 
they are robust in special regions of the incommensurate phase diagram. 
Orbital quantum fluctuations modify quantitatively spin-orbital order 
imposed by superexchange. In contrast, the spin-orbit coupling can lead 
to anisotropic spin and orbital patterns along the symmetry directions 
and cause a radical modification of the order imposed by the 
spin-orbital superexchange.
Our findings are expected to be of importance for future theoretical
understanding of experimental results for $4d$ transition metal oxides 
doped with $3d^3$ ions. We suggest how the local or global changes of 
the spin-orbital order induced by such impurities could be detected 
experimentally.
\end{abstract}


\maketitle

\section{Introduction}

The studies of strongly correlated electrons in transition metal
oxides (TMOs) focus traditionally on $3d$ materials \cite{Ima98},
mainly because of high-temperature superconductivity discovered in
cuprates and more recently in iron-pnictides, and because of colossal
magnetoresistance manganites.
The competition of different and complex types of order is ubiquitous
in strongly correlated TMOs mainly due to coupled spin-charge-orbital
where frustrated exchange competes with the kinetic energy of
charge carriers. The best known example is spin-charge competition
in cuprates, where spin, charge and superconducting orders intertwine
\cite{Ber11} and stripe order emerges in the normal phase as a
compromise between the magnetic and kinetic energy \cite{Lee06,Voj09}. 
Remarkable evolution of the stripe order under increasing doping is 
observed \cite{Yam98} and could be reproduced by the theory based on 
the extended Hubbard model \cite{Fle01}. Hole doping in cuprates 
corresponds to the removal of the spin degree of freedom. Similarly, 
hole doping in a simplest system with the orbital order in $d^1$ 
configuration removes locally orbital degrees of freedom and generates 
stripe phases which involve orbital polarons \cite{Wro10}. It was 
predicted recently that orbital domain walls in bilayer manganites 
should be partially charged as a result of competition between 
orbital-induced strain and Coulomb repulsion \cite{Li14}, which opens
a new route towards charge-orbital physics in TMOs. We will show 
below that the stripe-like order may also occur in doped spin-orbital 
systems. These systems are very challenging and their doping leads to 
very complex and yet unexplored spin-orbital-charge phenomena \cite{Zaa93}.

A prerequisite to the phenomena with spin-orbital-charge coupled
degrees of freedom is the understanding of undoped systems \cite{Tok00},
where the low-energy physics and spin-orbital order are dictated by
effective spin-orbital superexchange \cite{Kug82,Ole05,Woh11} and
compete with spin-orbital quantum fluctuations \cite{Fei97,Kha00,Kha05}.
Although ordered states occur in many cases, the most intriguing are 
quantum phases such as spin \cite{Bal10} or orbital \cite{Fei05} 
liquids. Recent experiments on a copper oxide Ba$_3$CuSb$_2$O$_9$ 
\cite{Nak12,Qui12} have triggered renewed efforts
in a fundamental search for a quantum spin-orbital liquid
\cite{Nor08,Karlo,Nas12,Mil14}, where spin-orbital order is absent 
and electron spins are randomly choosing orbitals which they occupy.
A signature of strong quantum effects in a spin-orbital system is a
disordered state which persists down to very low temperatures. A good
example of such a disordered spin-orbital liquid state is as well
FeSc$_2$S$_4$ which does not order in spite of finite Curie-Weiss
temperature $\Theta_{\rm CW}=-45$ K \cite{Fri04}, but shows instead 
signatures of quantum criticality \cite{Che09,Mit14}.

Spin-orbital interactions may be even more challenging ---
for instance previous attempts to find a spin-orbital liquid in the
Kugel-Khomskii model \cite{Fei97} or in LiNiO$_2$ \cite{Ver04} turned
out to be unsuccessful. In fact, in the former case certain types of 
exotic spin order arise as a consequence of frustrated and entangled
spin-orbital interactions \cite{Brz12,Brz13}, and a spin-orbital 
entangled resonating valence bond state was recently shown to be a
quantum superposition of strped spin-singlet covering on a square 
lattice \cite{Cza15}. In contrast,
spin and orbital superexchange have different energy scales and
orbital interactions in LiNiO$_2$ are much stronger and dominated by
frustration \cite{Rey01}. Hence the reasons behind the absence of
magnetic long range order are more subtle \cite{Rei05}.
In all these cases orbital fluctuations play a prominent role and
spin-orbital entanglement \cite{Ole12} determines the ground state.

The role of charge carriers in spin-orbital systems is under very
active investigation at present. In doped La$_{1-x}$(Sr,Ca)$_x$MnO$_3$
manganites several different types of magnetic order compete with one
another and occur at increasing hole doping \cite{Dag01,Dag05,Tok06}.
Undoped LaMnO$_3$ is an antiferromagnetic (AF) Mott insulator, with
large $S=2$ spins for $3d^4$ ionic configurations of Mn$^{3+}$ ions 
stabilized by Hund's exchange, coupled via the spin-orbital superexchange 
due to $e_g$ and $t_{2g}$ electron excitations \cite{Fei99}. The orbital 
$e_g$ degree of freedom is removed by hole doping when Mn$^{3+}$ ions 
are generated, and this requires careful modeling in the theory that 
takes into account both $3d^4$ and $3d^3$ electronic configurations of
Mn$^{3+}$ and Mn$^{4+}$ ions \cite{Kil99,Cuo02,Feh04,Dag04,Gec05,Kha11}.
In fact, the orbital order changes radically with increasing doping in
La$_{1-x}$(Sr,Ca)$_x$MnO$_3$ systems at the magnetic phase transitions
between different types of magnetic order \cite{Tok06}, as weel as at
La$_{0.7}$Ca$_{0.3}$MnO$_3$/BiFeO$_3$ heterostructures, where it offers 
a new route to enhancing multiferroic functionality \cite{She14}. The
double exchange mechanism \cite{deG60} triggers ferromagnetic (FM)
metallic phase at sufficient doping; in this phase the spin and orbital
degrees of freedom decouple and spin excitations are explained by the
orbital liquid \cite{KhaKi,Ole02}. Due to distinct magnetic and kinetic
energy scales, even low doping may suffice for a drastic change in the
magnetic order, as observed in electron-doped manganites \cite{Sak10}.

A rather unique example of a spin-orbital system with strongly
fluctuating orbitals, as predicted in the theory
\cite{Kha01,Kha04,Hor08} and seen experimentally
\cite{Ulr03,Zho07,Ree11}, are the perovskite vanadates with competing
spin-orbital order \cite{Fuj10}. In these $t_{2g}$ systems $xy$
orbitals are filled by one electron and orbital order of active
$\{yz,zx\}$ orbitals is strongly influenced by doping with Ca (Sr)
ions which replace Y (La) ones in YVO$_3$ (LaVO$_3$). In this case
finite spin-orbit coupling modifies the spin-orbital phase diagram
\cite{Hor03}. In addition, the AF order switches easily from the 
$G$-type AF ($G$-AF) to $C$-type AF ($C$-AF) order in the presence of
charge defects in Y$_{1-x}$Ca$_x$VO$_3$. Already at low $x\simeq 0.02$ 
doping the spin-orbital order changes and spectral weight is generated 
within the Mott-Hubbard gap \cite{Fuj08}. Although one might imagine 
that the orbital degree of freedom is thereby removed, a closer 
inspection shows that this is not the case as the orbitals are 
polarized by charge defects \cite{Hor11} and readjust near them 
\cite{Ave13}. Removing the orbital degree of freedom in vanadates 
would be only possible by electron doping generating instead $d^3$ 
ionic configurations, but such a doping by charge defects would be 
very different from the doping by transition metal ions of the same 
valence considered below.

Also in $4d$ materials spin-orbital physics plays a role \cite{Hot06},
as for instance in Ca$_{2-x}$Sr$_x$RuO$_4$ systems with Ru$^{4+}$ ions
in $4d^4$ configuration \cite{Miz01,Lee02,Kog04,Fan05,Sug13}.
Recently it has been shown that unconventional magnetism is possible
for Ru$^{4+}$ and similar ions where spin-orbit coupling plaus a role
\cite{Kha13,Kha14}.
Surprisingly, these systems are not similar to manganites but to
vanadates where one finds as well ions with active $t_{2g}$ orbitals.
In the case of ruthenates the $t_{2g}^4$ Ru$^{4+}$ ions have low $S=1$ 
spin as the splitting between the $t_{2g}$ and $e_g$ levels is large.
Thus the undoped Ca$_2$RuO$_4$ is a hole analogue of a vanadate
\cite{Kha01,Kha04}, with $t_{2g}$ orbital degree of freedom and 
$S=1$ spin per site in both cases. This gives new opportunities to
investigate spin-orbital entangled states in $t_{2g}$ system, observed
recently by angle resolved photoemission \cite{Vee14}.

Here we focus on a novel and very different doping from all those
considered above, namely on a \textit{substitutional} doping by other
magnetic ions in a plane built by transition metal and oxygen ions, for 
instance in the $(a,b)$ plane of a monolayer or in perovskite ruthenates
or vanadates. In this study we are interested primarily in doping of a
TMO with $t_{2g}$ orbital degrees of freedom, where doped magnetic ions
have no orbital degree of freedom and realize \textit{orbital dilution}.
In addition, we deal with the simpler case of $3d$ doped ions where we 
can neglect spin-orbit interaction which should not be ignored for $4d$
ions. We emphasize that in contrast to manganites where holes within 
$e_g$ orbitals participate in transport and are responsible for the 
colossal magnetoresitance, such doped hole are immobile due to the 
ionic potential at $3d$ sites and form defects in spin-orbital order 
of a Mott insulator. We encounter here a different situation from the 
dilution effects in the 2D $e_g$ orbital system considered so far 
\cite{Tan07} as we deel with magnetic ions at doped sites.
It is challenging to investigate how such impurities
modify locally or globally spin-orbital order of the host.

\begin{figure}[t!]
\includegraphics[width=.92\columnwidth]{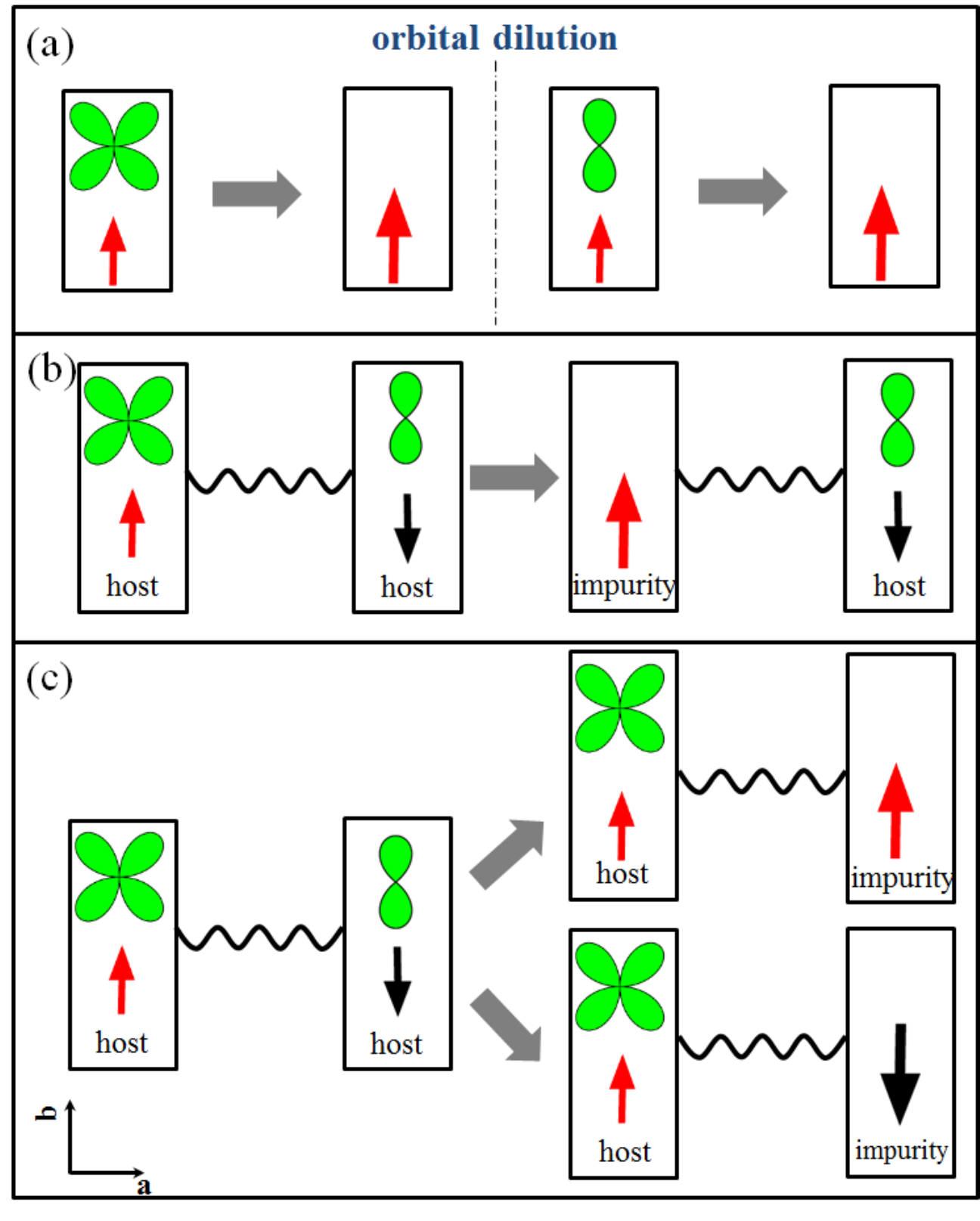}
\caption{
(a) Schematic view of the orbital \textit{dilution} when the $3d^3$
ion with no orbital degree of freedom and spin $S=3/2$ substitutes
$4d^4$ one with spin $S=1$ on a bond having specific spin and orbital
character in the host (gray arrows). Spins are shown by red arrows and
doubly occupied $t_{2g}$ orbitals (doublons) are shown by green
symbols for $a$ and $c$ orbitals, respectively.
(b) If an inactive orbital along the bond is removed by doping,
the total spin exchange is AF.
(c) On the contrary, active orbitals at the host site can lead
to either FM (top) or AF (bottom) exchange coupling, depending on the
energy levels mismatch and difference in the Coulomb couplings between
the impurity and the host.
We show the case when the host site is unchanged in the doping process.
\label{fig:bond_schem}}
\end{figure}

The doping which realizes this paradigm is by either Mn$^{4+}$ or
Cr$^{3+}$ ions with large $S=3/2$ spins stabilized by Hund's exchange,
and orbital dilution occurs either in a TMO with $d^2$ ionic
configuration as in the vanadium perovskites, or in $4d$ Mott
insulators as in ruthenates. It has been shown that dilute Cr
doping for Ru reduces the temperature of the orthorhombic distortion
and induces FM behavior in Ca$_2$Ru$_{1-x}$Cr$_x$O$_4$
(with $0<x<0.13$) \cite{Qi10}. It also induces surprising negative
volume thermal expansion via spin-orbital order.
Such defects, on one hand, can weaken the spin-orbital coupling in the
host, but on the other hand, may open a new channel of interaction
between the spin and orbital degree of freedom through the
host-impurity exchange, see Fig. \ref{fig:bond_schem}. Consequences
of such doping are yet unexplored and are expected to open a new route
in the research on strongly correlated oxides.

The physical example for the present theory are the insulating phases
of $3d-4d$ hybrid structures, where doping happens at $d^4$ transition
metal sites, and the value of the spin is locally changed from $S=1$
to $S=3/2$. As a demonstration of the highly nontrivial physics
emerging in $3d-4d$ oxides, remarkable effects have already been
observed, for instance, when Ru ions are replaced by Mn, Ti, Cr or
other $3d$ elements. The role of Mn doping in the SrRuO
Ruddlesden-Popper series is strongly linked to the dimensionality
through the number $n$ of RuO$_2$ layers in the unit cell. The Mn
doping of the SrRuO$_3$ cubic member drives the system from the
itinerant FM state to an insulating AF configuration in a continuous
way via a possible unconventional quantum phase transition \cite{Cao05}.
Doping by Mn ions in Sr$_{3}$Ru$_2$O$_{7}$ leads to a metal-to-insulator
transition and AF long-range order for more than $5\%$ Mn concentration
\cite{Hos12}. Subtle orbital rearrangement can occur at the Mn site,
as for instance the inversion of the crystal field in the $e_g$ sector
observed via x-ray absorption spectroscopy \cite{Hos08}. Neutron
scattering studies indicate the occurrence of an unusual $E$-type
antiferromagnetism in doped systems (planar order with FM zigzag chains
with AF order between them) with moments aligned along the $c$ axis
within a single bilayer~\cite{Mes12}.

\begin{figure}[t!]
\begin{center}
\includegraphics[width=.72\columnwidth]{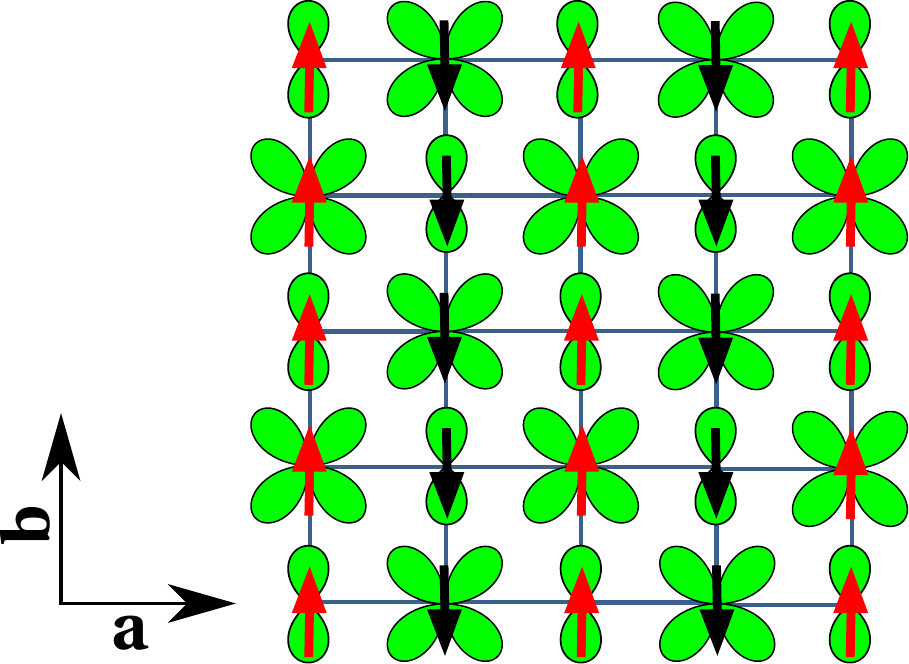}
\end{center}
\caption{
Schematic view of $C$-AF spin order coexisting with $G$-AO orbital 
order in the $(a,b)$ plane of an undoped Mott insulator with $4d^4$ 
ionic configurations. Spins
are shown by arrows while doubly occupied $xy$ and $yz$ orbitals ($c$
and $a$ doublons, see text) form a checkerboard pattern. Equivalent
spin-orbital order is realized for V$^{3+}$ ions in $(b,c)$ planes of
LaVO$_3$ \cite{Fuj10}, with orbitals standing for empty orbitals 
(holes).}
\label{fig:host}
\end{figure}

Furthermore, the more extended $4d$ orbitals would \textit{a priori}
suggest a weaker correlation than in $3d$ TMOs due to a reduced ratio
between the intraatomic Coulomb interaction and the electron bandwidth.
Nevertheless, the (effective) $d$-bandwidth is reduced by the changes
in the $3d$-$2p$-$3d$ bond angles in distorted structures which
typically arise in these materials. This brings these systems on the 
verge of a metal-insulator transition \cite{Kim09}, or even into the 
Mott insulating state with spin-orbital order, see Fig. \ref{fig:host}.
Hence, not only $4d$ materials share common features with $3d$ systems,
but are also richer due to their sensitivity to the lattice structure
and to relativistic effects due to larger spin-orbit \cite{Ann06}
or other magneto-crystalline couplings.

To simplify the analysis we assume that onsite Coulomb interactions are
so strong that charge degrees of freedom are projected out, and only
virtual charge transfer can occur between $3d$ and $4d$ ions via the
oxygen ligands. For convenience, we define the orbital degree of
freedom as a \textit{doublon} (double occupancy) in the $t_{2g}^4$
configuration. The above $3d$ doping leads then
effectively to the removal of a doublon in one of $t_{2g}$ orbitals
which we label as $\{a,b,c\}$ (this notation is introduced in Ref.
\cite{Kha05} and explained below) and to replacing it by a $t_{2g}^3$
ion. To our knowledge, this is the only example of removing the
orbital degree of freedom in $t_{2g}$ manifold realized so far and
below we investigate possible consequences of this phenomenon.
Another possibility of orbital dilution which awaits experimental
realization would occur when a $t_{2g}$ degree of freedom is removed 
by replacing a $d^2$ ion by a $d^3$ one, as for instance by Cr$^{3+}$ 
doping in a vanadate --- here a doublon is an empty $t_{2g}$ orbital, 
i.e., filled by two holes.

Before presenting the details of the quantitative analysis, let us
concentrate of the main idea of the superexchange modified by doping 
in a spin-orbital system. The $d^3$ ions have singly occupied all 
three $t_{2g}$ orbitals and $S=3/2$ spins due to Hund's exchange.
While a pair of $d^3$ ions, e.g. in SrMnO$_3$, is coupled by AF
superexchange \cite{Ole02}, the superexchange for the $d^3-d^4$ bond
has a rather rich structure and may also be FM. The spin exchange
depends then on whether the orbital degree of freedom is active and
participates in charge excitations along a considered bond or electrons
of the doublon cannot move along this bond due to the symmetry of
$t_{2g}$ orbital, as explained in Fig. \ref{fig:bond_schem}. This
qualitative difference to systems without active orbital degrees of
freedom is investigated in detail in Sec. \ref{sec:model}.

The main outcomes of our analysis are:
(i) the determination of the effective spin-orbital exchange Hamiltonian
describing the low-energy sector for the $3d-4d$ hybrid structure,
(ii) establishing that a $3d^3$ impurity without an orbital degree
of freedom modifies the orbital order in the $4d^4$ host,
(iii) providing the detailed way how the microscopic spin-orbital order
within the $4d^4$ host is modified around the $3d^3$ impurity, and
(iv) suggesting possible spin-orbital patterns that arise due to  
periodic and finite substitution (doping) of $4d$ atoms in the host 
by $3d$ ones.
The emerging physical scenario is that the $3d$ impurity acts as
an orbital vacancy when the host-impurity coupling is weak and as an
orbital polarizer of the bonds active $t_{2g}$ doublon configurations
when it is strongly coupled to the host. The tendency to polarize the
host orbitals around the impurity turns out to be robust and independent 
of spin configuration. Otherwise, it is the resulting orbital
arrangement around the impurity and the strength of Hund's coupling at
the impurity that set the character of the host-impurity magnetic
exchange.

The remaining of the paper is organized as follows. In Sec.
\ref{sec:model} we introduce the effective model describing the
spin-orbital superexchange at the $3d-4d$ bonds which serves to
investigate the changes of spin and orbital order around individual
impurities and at finite doping. We arrive at a rather general
formulation which emphasizes the impurity orbital degree of freedom,
being a doublon, and present some technical details of the derivation
in Appendix A. The strategy we adopt is to analyze first the ground
state properties of a single $3d^3$ impurity surrounded by $4d^4$ atoms
by investigating how the spin-orbital pattern in the host may be
modified at the nearest neighbor (NN) sites to the $3d$ atom. This
study is performed for different spin-orbital patterns of the $4d$ host
with special emphasis on the alternating FM chains ($C$-AF order) which
coexist with $G$-type alternating orbital ($G$-AO) order, see Fig. 
\ref{fig:host}. We address the impurity problem within the classical 
approximation in Sec. \ref{sec:mfa}. As explained in Sec. \ref{sec:two}, 
there are two nonequivalent cases which depend on the precise 
modification of the orbital order by the $3d$ impurity, doped either 
to replace a doublon in $a$ orbital (Sec. \ref{sec:impa}) or the one 
in $c$ orbital (Sec. \ref{sec:impc}).

Starting from the single impurity solution we next address periodic
arrangements of $3d$ atoms at different concentrations. We demonstrate
that the spin-orbital order in the host can be radically changed by
the presence of impurities, leading to striped patterns with
alternating FM/AF domains and islands of fully FM states. In Sec.
\ref{sec:dopg} we consider the modifications of spin-orbital order 
which arise at periodic doping with macroscopic concentration. 
Here we limit ourselves to two representative cases:
(i) commensurate $x=1/8$ doping in Sec. \ref{sec:dop8}, and
(ii) two doping levels $x=1/5$ and $x=1/9$ being incommensurate with 
underlying two-sublattice order (Fig. \ref{fig:host}) which implies 
simultaneous doping at two sublattices, i.e., at both $a$ and $c$ 
doublon sites, as presented in Secs. \ref{sec:dop5} and \ref{sec:dop9}.
Finally, in Sec. \ref{sec:orb} we investigate the modifications of the
classical phase diagram induced by quantum fluctuations, and in Secs.
\ref{sec:soc} and \ref{sec:JH} we discuss representative results
obtained for finite spin-orbit coupling (calculation details of the
treatment of spin-orbit interaction are presented in Appendix B).
The paper is concluded by a general discussion of possible emerging
scenarios for the $3d^3$ impurities in $4d^4$ host, a summary of
the main results and perspective of future experimental
investigations of orbital dilution in Sec. \ref{sec:sum}.

\section{The spin-orbital model}
\label{sec:model}

In this Section we consider a $3d$ impurity in a strongly correlated
$4d$ TMO and derive the effective $3d^3-4d^4$ spin-orbital 
superexchange. It follows from the coupling between $3d$ and $4d$
orbitals via oxygen $2p$ orbitals due to the $p-d$ hybridization. 
In a strongly correlated
system it suffices to concentrate on a pair of atoms forming a bond
$\langle ij\rangle$, as the effective interactions are generated by
charge excitations $d^4_id^4_j\leftrightharpoons d^5_id^3_j$ along a
single bond \cite{Ole05}. In the reference $4d$ host both atoms on
the bond $\langle ij\rangle$ are equivalent and one considers,
\begin{equation}
H(i,j)=H_t(i,j)+H_{\rm int}(i)+H_{\rm int}(j).
\label{host}
\end{equation}
The Coulomb interaction $H_{\rm int}(i)$ is local at site $i$ and we
describe it by the degenerate Hubbard model \cite{Ole83}, see below.

We implement a strict rule that the hopping within the $t_{2g}$ sector
is allowed in a TMO only between two neighboring orbitals of the same
symmetry which are active along the bond direction
\cite{Kha00,Har03,Dag08}, and neglect the interorbital processes
originating from the octahedral distortions such as rotation or
tilting. Indeed, in ideal undistorted (perovskite or square lattice)
geometry the orbital flavor is conserved as long as the spin-orbit
coupling may be neglected. The interorbital hopping elements are
smaller by at least one order of magnitude and may be treated as
corrections in cases where distortions play a role to the overall 
scenario established below.

The kinetic energy for a representative $3d$-$2p$-$4d$ bond, i.e.,
after projecting out the oxygen degrees of freedom, is given by
the hopping in the host $\propto t_h$ between sites $i$ and $j$,
\begin{equation}
H_t(i,j)=-t_h\sum_{\mu(\gamma),\sigma}
\left(d_{i\mu\sigma}^{\dagger}d_{j\mu\sigma}^{}
+d_{j\mu\sigma}^{\dagger}d_{i\mu\sigma}^{}\right).
\label{eq:hop4d}
\end{equation}
Here $d_{i\mu\sigma}^{\dagger}$ are the electron creation operators
at site $i$ in the spin-orbital state $(\mu\sigma)$. The bond
$\langle ij\rangle$ points along one of the two crystallographic
directions, $\gamma=a,b$, in the two-dimensional (2D) square lattice. 
Without distortions, only two out of three $t_{2g}$ orbitals are active 
along each bond $\langle 12\rangle$ and contribute to $H_t(i,j)$, while 
the third orbital lies in the plane perpendicular to the $\gamma$ axis 
and thus the hopping via oxygen is forbidden by symmetry. This motivates 
a convenient notation as follows \cite{Kha00},
\begin{equation}
\left|a\right\rangle\equiv\left|yz\right\rangle, \quad
\left|b\right\rangle\equiv\left|xz\right\rangle, \quad
\left|c\right\rangle\equiv\left|xy\right\rangle,
\label{eq:or_defs}
\end{equation}
with the $t_{2g}$ orbital inactive along a given direction
$\gamma\in\{a,b,c\}$ labeled by the index $\gamma$. We consider a
2D square lattice with transition metal ions
connected via oxygen orbitals as in a RuO$_2$ $(a,b)$ plane of
Ca$_2$RuO$_4$ (SrRuO$_3$). In this case $|a\rangle$ ($|b\rangle$)
orbitals are active along the $b$ ($a$) axis, while $|c\rangle$
orbitals are active along both $a,b$ axes.

To derive the superexchange in a Mott insulator, it is sufficient to
consider a bond which connects nearest neighbor sites,
$\langle ij\rangle\equiv\langle 12\rangle$. Below we consider a bond
between an impurity site $i=1$ occupied by a $3d$ ion and a neighboring
host $4d$ ion at site $j=2$. The Hamiltonian for this bond can be then
expressed in the following form,
\begin{equation}
H(1,2)=H_t(1,2)+H_{\rm int}(1)+H_{\rm int}(2)+H_{\rm ion}(2).
\label{Hub}
\end{equation}
The total Hamiltonian contains the kinetic energy term $H_t(1,2)$ 
describing the electron charge transfer via oxygen orbitals, the onsite 
interaction terms $H_{\rm int}(m)$ for the $3d$ $(4d)$ ion at site $m=1,2$,
and the local potential of the $4d$ atom, $H_{ion}(2)$, which takes into
account the mismatch of the energy level structure between the two ($4d$
and $3d$) atomic species and prevents valence fluctuations when the host
is doped, even in the absence of local Coulomb interaction.

The kinetic energy in Eq. (\ref{Hub}) is given by,
\begin{equation}
H_t(1,2)=-t\sum_{\mu(\gamma),\sigma}
\left(d_{1\mu\sigma}^{\dagger}d_{2\mu\sigma}^{}
+d_{2\mu\sigma}^{\dagger}d_{1\mu\sigma}^{}\right),
\label{eq:hop}
\end{equation}
where $d_{m\mu\sigma}^{\dagger}$ is the electron creation operator at
site $m=1,2$ in the spin-orbital state $(\mu\sigma)$. The bond
$\langle 12\rangle$ points along one of the two crystallographic
directions, $\gamma=a,b$, and again the orbital flavor is conserved
\cite{Kha00,Har03,Dag08}.

The Coulomb interaction on an atom at site $m=1,2$ depends on two
parameters \cite{Ole83}:
(i) intraorbital Coulomb repulsion $U_m$, and
(ii) Hund's exchange $J_m^H$.
The label $m$ stands for the ion and distinguishes between these
terms at the $3d$ and $4d$ ion, respectively. The interaction is
expressed in the form,
\begin{eqnarray}
H_{\rm int}(m)&=& U_m\sum_{\mu}n_{m\mu\uparrow}n_{m\mu\downarrow}
- 2J_m^{H}\sum_{\mu<\nu}\vec{S}_{m\mu}\!\cdot\!\vec{S}_{m\nu}
\nonumber \\
&+&\left(U_m-\frac{5}{2}J_m^{H}\right)
\sum_{{\mu<\nu\atop \sigma\sigma'}}n_{m\mu\sigma}n_{m\nu\sigma'}
\nonumber \\
&+&   J_m^{H}\sum_{\mu\not=\nu}d_{m\mu\uparrow}^{\dagger}
d_{m\mu\downarrow}^{\dagger}d_{m\nu\downarrow}^{}d_{m\nu\uparrow}^{}.
\label{Hint}
\end{eqnarray}
The terms standing in the first line of Eq. (\ref{Hint})
contribute to the magnetic instabilities in degenerate Hubbard
model \cite{Ole83} and decide about spin order, both in an
itinerant system and in a Mott insulator. The remaining terms
contribute to the multiplet structure and are of importance for
the correct derivation of the superexchange which follows from
charge excitations, see below.

Finally, we include a local potential on the $4d$ atom which encodes
the energy mismatch between the host and the impurity orbitals close
to the Fermi level and prevents valence fluctuations on the $4d$ ion
due to the $3d$ doping. This term has the following general structure,
\begin{equation}
H_{\rm ion}(2)=I_2^e\left(4-\sum_{\mu,\sigma}n_{2\mu\sigma}\right)^2,
\label{eq:ion2}
\end{equation}
with $\mu=a,b,c$.

\begin{figure}[t]
\includegraphics[clip,width=0.6\columnwidth]{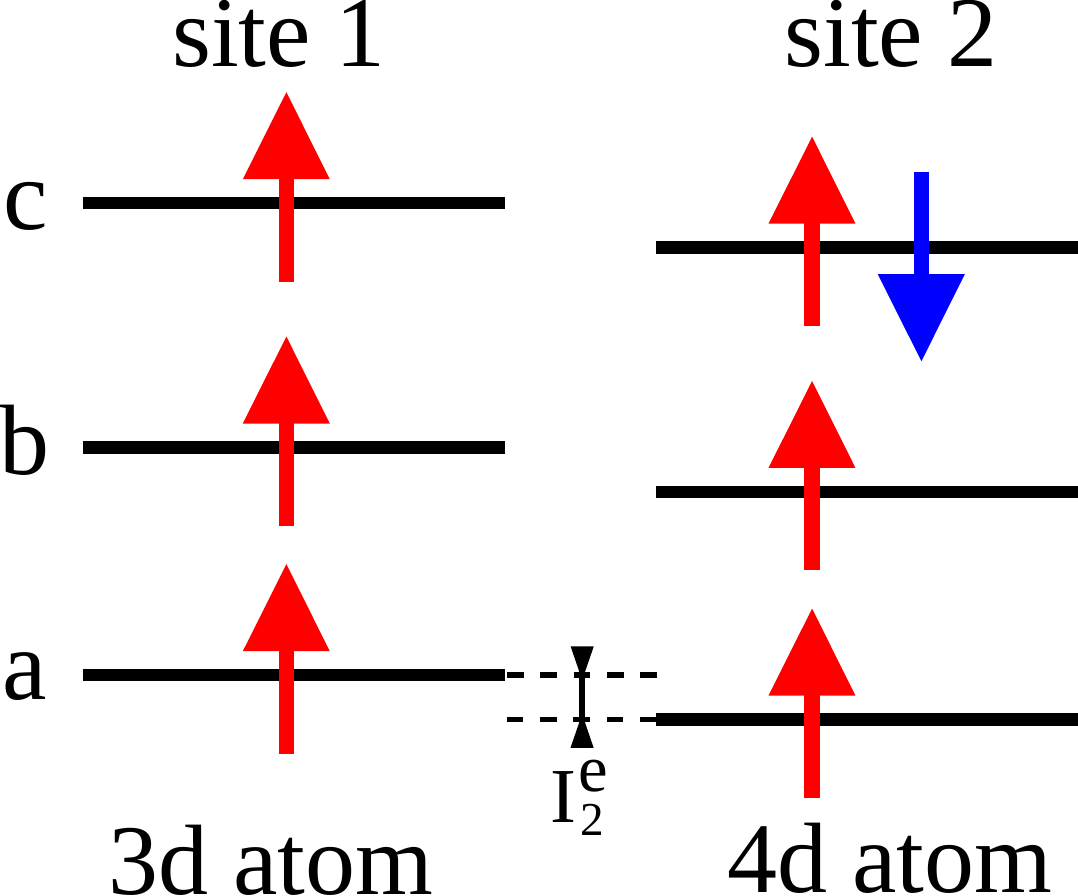}
\caption{
Schematic representation of one configuration belonging to the manifold
of 36 degenerate ground states for a representative $3d-4d$ bond
$\langle 12\rangle$ as given by the local Coulomb Hamiltonian
$H_{\rm int}(m)$ (\ref{Hint}) with $m=1,2$. The dominant
exchange processes considered here are those that move one of the four
electrons on the $4d$ atom to the $3d$ neighbor and back. The stability
of the $3d^{3}$-$4d^{4}$ charge configurations is provided by the local
potential energy $I_{2}^{e}$, see Eq. (\ref{eq:ion2}). }
\label{fig:3d4d}
\end{figure}

The effective Hamiltonian for the low energy processes is derived from
$H(1,2)$ (\ref{Hub}) by a second order expansion for charge excitations
generated by $H_t(1,2)$, and treating the remaining part of $H(1,2)$ as 
an unperturbed Hamiltonian. We are basically interested in virtual
charge excitations in the manifold of degenerate ground states of a
pair of $3d$ and $4d$ atoms on a bond, see Fig. \ref{fig:3d4d}.
These quantum states are labeled as $\left\{e_1^k\right\}$ with
$k=1,\dots,4$ and $\left\{e_{2}^p\right\}$ with $p=1,\dots,9$ and their
number follows from the solution of the onsite quantum problem for the
Hamiltonian $H_{\rm int}(i)$. For the $3d$ atom the relevant states can 
be classified according to the four components of the total spin 
$S_1=3/2$ for the $3d$ impurity atom at site $m=1$, three components of 
$S_2=1$ spin for the $4d$ host atom at site $m=2$ and for the three 
different positions of the double occupied orbital (doublon).
Thus, the effective Hamiltonian will contain spin products
$(\vec{S}_1\!\cdot\!\vec{S}_2)$ between spin operators defined as,
\begin{equation}
\vec{S}_{m}=\frac{1}{2}\sum_{\gamma} d_{m\gamma\alpha}^{\dagger}
\vec{\sigma}_{\alpha\beta}^{}d_{m\gamma\beta}^{},
\label{eq:spin_op}
\end{equation}
for $m=1,2$ sites and the operator of the doublon position at site
$m=2$,
\begin{equation}
D_2^{(\gamma)}=\left(
d_{2\gamma\uparrow}^{\dagger}d_{2\gamma\uparrow}^{}\right)
\left(d_{2\gamma\downarrow}^{\dagger}d_{2\gamma\downarrow}^{}\right).
\label{eq:dub}
\end{equation}
The doublon operator identifies the orbital $\gamma$ within the
$t_{2g}$ manifold of the $4d$ ion with a double occupancy (occupied by
the doublon) and stands in what follows for the orbital degree of
freedom. It is worth noting that the hopping (\ref{eq:hop}) does not
change the orbital flavor thus we expect that the resulting Hamiltonian
is diagonal in the orbital degrees of freedom with only $D_2^{(\gamma)}$
operators.

Following the standard second order perturbation expansion for
spin-orbital systems \cite{Ole05}, we can write the matrix elements of
the low energy exchange Hamiltonian, ${\cal H}_J^{(\gamma)}(i,j)$, for
a bond $\langle 12\rangle\parallel\gamma$ along the $\gamma$ axis as
follows,
\begin{eqnarray}
&&\!\big\langle e_{1}^{k},e_{2}^{l}\big|{\cal H}_J^{(\gamma)}(1,2)
\big|e_{1}^{k'},e_{2}^{l'}\big\rangle  =
-\sum_{n_{1},n_{2}}\frac{1}{\varepsilon_{n1}+\varepsilon_{n2}}
\nonumber \\
\!\times&\!\big\langle& e_{1}^{k},e_{2}^{l}\big|H_t(1,2)\big|
n_{1},n_{2}\big\rangle\!  \times
\!\big\langle n_{1},n_{2}\big|H_t(1,2)\big|
e_{1}^{k'}\!,e_{2}^{l'}\!\big\rangle,
\label{eq:pert_exp}
\end{eqnarray}
with $\varepsilon_{nm}=E_{n,m}-E_{0,m}$ being the excitation energies
for atoms at site $m=1,2$ with respect to the unperturbed ground state.
The superexchange Hamiltonian ${\cal H}_J^{(\gamma)}(1,2)$ for a bond
along $\gamma$ can be expressed in a matrix form by a $36\times 36$
matrix, with dependence on $U_m$, $J_m^H$, and $I_{e}$ elements.
There are two types of charge excitations:
(i)~$d^3_1d^4_2\leftrightharpoons d^4_1d^3_2$ one which creates a
doublon at the $3d$ impurity, and
(ii)~$d^3_1d^4_2\leftrightharpoons d^2_1d^5_2$ one which adds another
doublon at the $4d$ host site in the intermediate state.
The second type of excitations involves more doubly occupied orbitals
and has much larger excitation energy. It is therefore only a small
correction to the leading term (i), as we discuss in Appendix A.

Similar as in the case of doped manganites \cite{Ole02}, the dominant
contribution to the effective low-energy spin-orbital Hamiltonian for
the $3d-4d$ bond stems from the
$d^3_1d^4_2\leftrightharpoons d^4_1d^3_2$ charge excitations, as they
do not involve an extra double occupancy and the Coulomb energy $U_2$.
The $3d^3_14d^4_2\leftrightharpoons 3d^4_14d^3_2$ charge excitations
can be analyzed in a similar way as the
$3d^3_i3d^4_j\leftrightharpoons 3d^4_i3d^3_j$ ones for an
$\langle ij\rangle$ bond in doped manganites \cite{Ole02}.
In both cases the total number of doubly occupied orbitals does not
change, so the main contributions come due to Hund's exchange.
In the present case, one more parameter plays a role,
\begin{equation}
\Delta=I_{e}+3(U_{1}-U_{2})-4(J_{1}^{H}-J_{2}^{H}),
\label{Delta}
\end{equation}
which stands for the mismatch potential energy (\ref{eq:ion2})
renormalized by the onsite Coulomb interactions $\{U_m\}$ and by Hund's 
exchange $\{J_m^H\}$. On a general ground we expect $\Delta$ to be a
positive quantity, since the repulsion $U_m$ should be larger for
smaller $3d$ shells than for the $4d$ ones and $U_m$ is the
largest energy scale in the problem.

Let us have a closer view on this dominant contribution of the
effective low-energy spin-orbital Hamiltonian for the $3d-4d$ bond,
given by Eq. (\ref{eq:s-ex}).
For the analysis performed below and the clarity of our presentation
it is convenient to introduce some scaled parameters related to the
interactions within the host and between the host and the impurity.
For this purpose we employ the exchange couplings $J_{\rm imp}$ and
$J_{\rm host}$,
\begin{eqnarray}
\label{eq:bothJ}
J_{\rm imp}&=&\frac{t^{2}}{4\Delta}, \\
J\label{eq:hostJ}
_{\rm host}&=&\frac{4t_h^{2}}{U_2},
\end{eqnarray}
which follow from the virtual charge excitations generated by the
kinetic energy, see Eqs. (\ref{eq:hop4d}) and (\ref{eq:hop}). We
use their ratio to investigate the influence of the impurity on the
spin-orbital order in the host. Here $t_h$ is the hopping amplitude
between two $t_{2g}$ orbitals at NN $4d$ atoms, $J_{2}^{H}$ and
$U_{2}$ refer to the host, and $\Delta$ (\ref{Delta}) is the
renormalized ionization energy of the $3d-4d$ bonds. The results
depend as well on Hund's exchange element for the impurity and on
the one at host atoms,
\begin{eqnarray}
\label{eq:etai}
\eta_{\rm imp} &=&\frac{J_{1}^{H}}{\Delta},\\
\label{eq:etah}
\eta_{\rm host}&=&\frac{J_2^{H}}{U_2},
\end{eqnarray}
Note that the ratio introduced for the impurity, $\eta_{\rm imp}$
(\ref{eq:etai}), has here a different meaning from Hund's exchange
used here for the host, $\eta_{\rm host}$ (\ref{eq:etah}), which
cannot be too large by construction, i.e., $\eta_{\rm host}<1/3$.

\begin{figure}[t!]
\begin{center}
\includegraphics[clip,width=0.9\columnwidth]{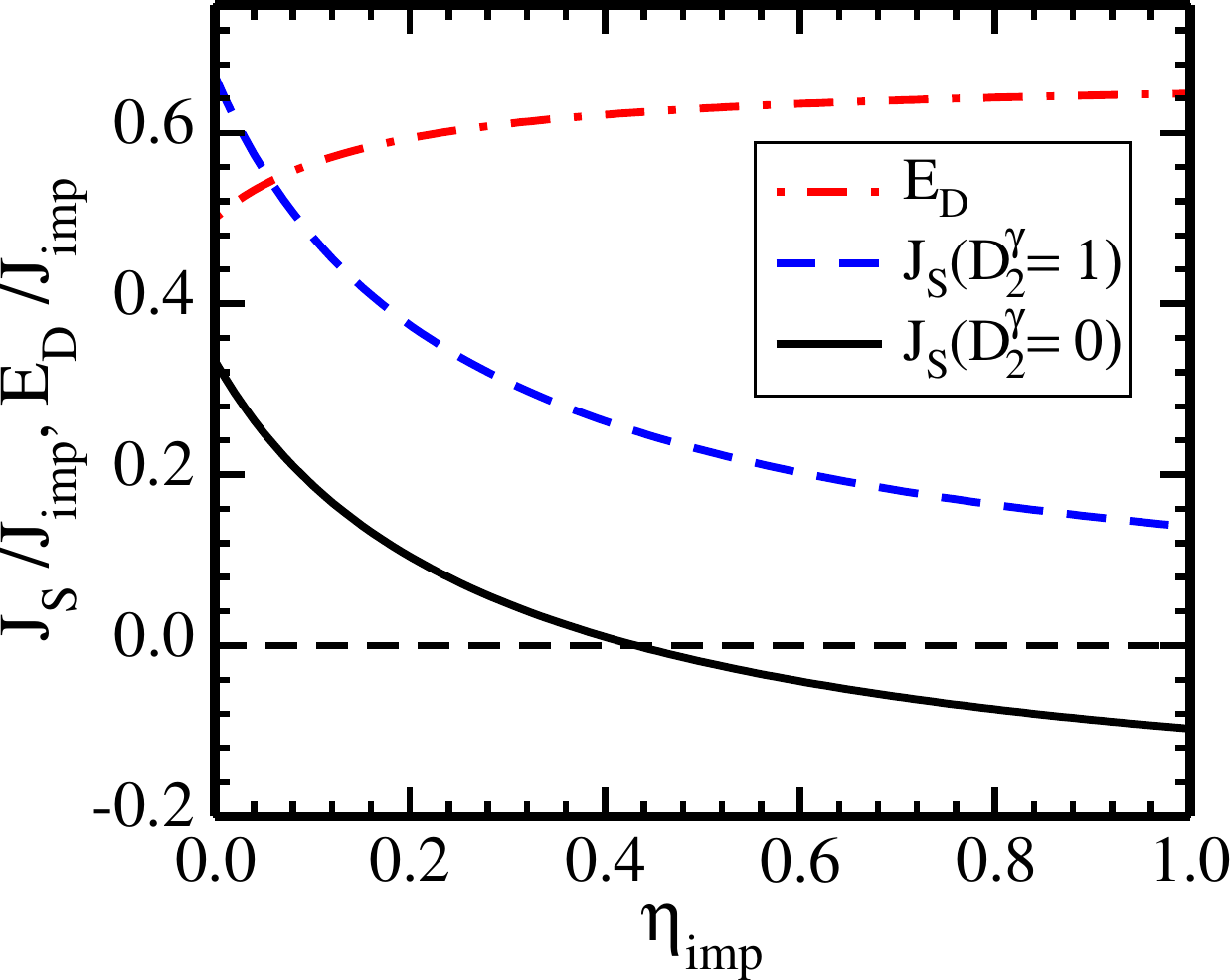}
\end{center}
\caption{
Evolution of the spin exchange $J_S(D_2^{(\gamma)})$ and the doublon 
energy $E_D$, both given in Eq. (\ref{eq:H123}) for increasing 
Hund's exchange $\eta_{\rm imp}$ at the impurity.
\label{fig:JSEd}}
\end{figure}

With the parametrization introduced above, the dominant term in the
impurity-host Hamiltonian for the impurity spin $\vec{S}_i$ interacting
with the neighboring host spins $\{\vec{S}_j\}$ at $j\in{\cal N}(i)$,
deduced from ${\cal H}_{3d-4d}^{(\gamma)}(1,2)$ Eq. (\ref{eq:s-ex}),
can be written in a rather compact form as follows
\begin{equation}
{\cal H}_{3d-4d}(i)\simeq\sum_{\gamma,j\in{\cal N}(i)}
\left\{J_{S}(D_j^{(\gamma)})(\vec{S}_i\!\cdot\!\vec{S}_j)
+ E_D D_j^{(\gamma)}\right\},
\label{eq:H123}
\end{equation}
where the orbital (doublon) dependent spin couplings
$J_S(D_j^{(\gamma)})$ and the doublon energy $E_D$
depend on $\eta_{\rm imp}$. The evolution of the exchange couplings are
shown in Fig. \ref{fig:JSEd}. We note that the dominant energy scale is
$E_D^{\gamma}$, so for a single $3d-4d$ bond the doublon will avoid
occupying the inactive ($\gamma$) orbital and the spins will couple
with $J_{S}(D_j^{(\gamma)}=0)$ which can be either AF if
$\eta_{\rm imp}\lesssim0.43$ or FM if $\eta_{\rm imp}>0.43$.
Thus the spins at $\eta_{\rm imp}=\eta_{\rm imp}^c\simeq 0.43$ will
decouple according to the ${\cal H}_J^{(\gamma)}(i,j)$ exchange.

Let us conclude this Section by writing the
complete superexchange Hamiltonian,
\begin{equation}
{\cal H}={\cal H}_{3d-4d}+{\cal H}_{4d-4d}+{\cal H}_{so},
\label{fullH}
\end{equation}
where ${\cal H}_{3d-4d}\equiv\sum_i{\cal H}_{3d-4d}(i)$ includes all
the $3d-4d$ bonds around impurities, ${\cal H}_{4d-4d}$ stands for the
the effective spin-orbital Hamiltonian for the $4d$ host bonds, and
${\cal H}_{so}$ is the spin-orbit interaction in the host. The former
term we explain below, while the latter one is defined in Sec. 
\ref{sec:soc}, where we analyze the quantum corrections and the 
consequences of spin-orbit interaction. The superexchange in the
host for the bonds $\langle ij\rangle$ along the $\gamma=a,b$ axes
\cite{Cuo06},
\begin{equation}
{\cal H}_{4d-4d}=J_{\rm host}\sum_{\langle ij\rangle\parallel\gamma}
\left\{J_{ij}^{(\gamma)}(\vec{S}_{i}\!\cdot\!\vec{S}_{j}+1) 
+K_{ij}^{(\gamma)}\right\},
\label{eq:Hhost}
\end{equation}
depends on $J_{ij}^{(\gamma)}$ and $K_{ij}^{(\gamma)}$ operators acting
only in the orbital space. They are expressed in terms of the
pseudospin operators defined in the orbital subspace spanned by the two 
orbital flavors active along a given direction $\gamma$, i.e.,
\begin{eqnarray}
J_{ij}^{(\gamma)} & = & \frac{1}{2}\left(2r_{1}+1\right)
\left(\vec{\tau}_{i}^{}\!\cdot\!\vec{\tau}_{j}^{}\right)^{(\gamma)}
-\frac{1}{2}r_2\left(\tau_i^z\tau_j^z\right)^{(\gamma)}
\nonumber \\
& + & \frac{1}{8}\left(n_{i}^{}n_{j}^{}\right)^{(\gamma)}
\left(2r_{1}\!-\!r_{2}\!+\!1\right)-\frac{1}{4}r_1
\left(n_{i}^{}+n_{j}^{}\right)^{(\gamma)}, \\
K_{ij}^{(\gamma)} & = &
r_1\left(\vec{\tau}_{i}\!\cdot\!\vec{\tau}_{j}\right)^{(\gamma)}
+\!r_2\left(\tau_i^z\tau_j^z\right)^{(\gamma)}
+\!\frac14\left(r_1+r_2\right)\left(n_{i}^{}n_{j}^{}\right)^{(\gamma)}
\nonumber \\
& - & \frac14\left(r_1+1\right)\left(n_i^{}+n_j^{}\right)^{(\gamma)}.
\end{eqnarray}
with
\begin{equation}
r_1=\frac{\eta_{\rm host}}{1-3\eta_{\rm host}}, \hskip .5cm
r_2=\frac{\eta_{\rm host}}{1+2\eta_{\rm host}},
\label{rr}
\end{equation}
standing for the multiplet structure in charge excitations,
and the orbital operators $\{\vec{\tau}_i^{\,(\gamma)},n_i^{(\gamma)}\}$
that for the $\gamma=c$ axis take the form:
\begin{eqnarray}
\vec{\tau}_{i}^{\,(c)}&=&\frac{1}{2}\big(\begin{array}[t]{cc}
a_{i}^{\dagger} & b_{i}^{\dagger}\end{array}\big)\cdot\vec{\sigma}\cdot
\big(\begin{array}[t]{cc}a_i^{} & b_i^{}\end{array}\big)^{\intercal}, \\
n_{i}^{(c)}&=&a_{i}^{\dagger}a_{i}^{}+b_{i}^{\dagger}b_{i}^{}.
\end{eqnarray}
For the directions $\gamma=a,b$ in the considered $(a,b)$ plane one
finds equivalent expressions by cyclic permutation of the axis labels
$\{a,b,c\}$ in the above formulas. This problem is isomorphic with
the spin-orbital superexchange in the vanadium perovskites
\cite{Kha01,Kha04}, where a hole in the $\{a,b\}$ doublet plays an
equivalent role to the doublon in the present case.
The operators $\{a_{i}^{\dagger},b_{i}^{\dagger},c_{i}^{\dagger}\}$ are
the doublon (hard core boson) creation operators in the orbital
$\gamma=a,b,c$, respectively, and they satisfy the local constraint,
\begin{equation}
a_{i}^{\dagger}a_i^{}+b_{i}^{\dagger}b_i^{}+c_{i}^{\dagger}c_i^{}=1,
\end{equation}
meaning that exactly \textit{one} doublon (\ref{eq:dub}) occupies one
of the three $t_{2g}$ orbitals at each site $i$. These bosonic
occupation operators coincide with the previously used doublon
occupation operators $D^{(\gamma)}_j$, i.e.,
$D^{(\gamma)}_j=\gamma^{\dagger}_j\gamma_j$ with $\gamma=a,b,c$.
Below we follow first the classical procedure to determine the ground 
states of single impurities in Sec. \ref{sec:imp}, and at macroscopic 
doping in Sec \ref{sec:dop}.

\section{Single 3\MakeLowercase{d} impurity in 4\MakeLowercase{d} host}
\label{sec:imp}

\subsection{Classical treatment of the impurity problem}
\label{sec:mfa}

\begin{figure}[t!]
\begin{center}
\includegraphics[width=0.7\columnwidth]{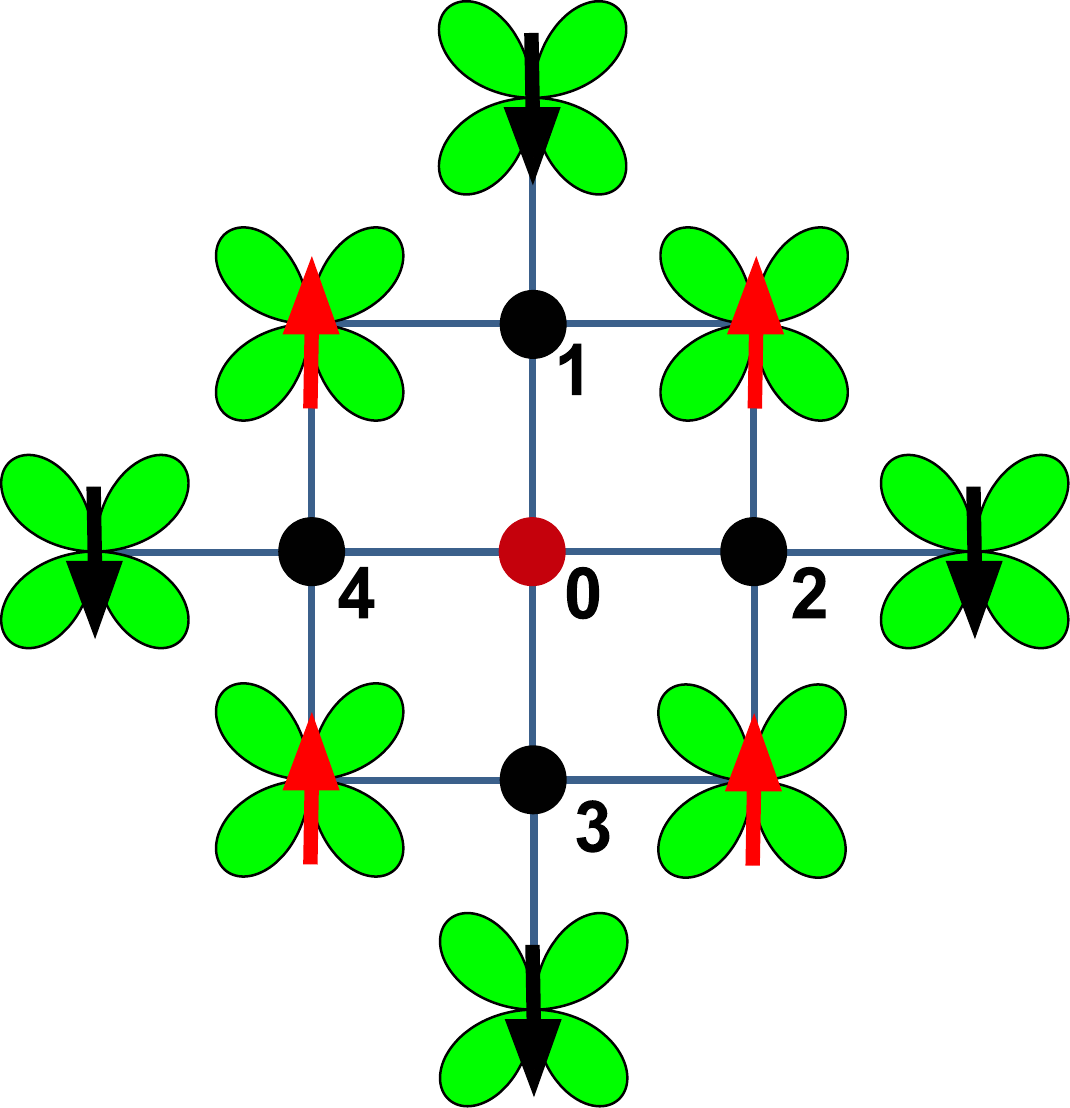}
\end{center}
\protect\caption{Schematic top view of the cluster used to obtain the
phase diagrams of the $3d$ impurity within the $4d$ host in an $(a,b)$
plane. The impurity is at the central site $i=0$ which belongs to the 
$c$ orbital host sublattice. For the outer sites in this cluster the 
spin-orbital configuration is fixed and determined by the undoped $4d$ 
host (with spins and $c$ orbitals shown here) having $C$-AF/$G$-AO 
order, see Fig. \ref{fig:host}.
For the central $i=0$ site the spin state and for the host sites
$i=1,\dots,4$ the spin-orbital configurations are
determined by minimizing the energy of the cluster.
}
\label{fig:setup}
\end{figure}

In this Section we describe the methodology that we applied for the
determination of the phase diagrams for a single impurity reported
below in Secs. \ref{sec:impa}, and next at macroscopic
doping, as presented in Sec. \ref{sec:dop}. Let us consider first
the case of a single $3d$ impurity in the $4d$ host. Since the
interactions in the model Hamiltonian are only effective ones between
NN sites, it is sufficient to study the modification
of the spin-orbital order around the impurity for a given spin-orbital
configuration of the host by investigating a cluster of $L=13$ sites
shown in Fig.~\ref{fig:setup}. We assume the $C$-AF spin order
(FM chains coupled antiferromagnetically) accompanied by $G$-AO order
within the host which is the spin-orbital order occurring for the
realistic parameters of a RuO$_2$ plane \cite{Cuo06}, see 
Fig.~\ref{fig:host}. Such a spin-orbital pattern turns out to be the 
most relevant one when considering the competition between the host and
the impurity as due to the AO order within the $(a,b)$ plane. Other
possible configurations with uniform orbital order and AF spin
pattern, e.g. $G$-AF order, will also be considered in the discussion
throughout the manuscript. The sites $i=1,2,3,4$ inside the cluster in 
Fig. \ref{fig:setup} have active spin and orbital degrees of freedom
while the impurity at site $i=0$ has only spin degree of freedom.
At the remaining sites the spin-orbital configuration is assigned,
following the order in the host, and it does not change along the
computation.

To determine the ground state we assume that the spin-orbital degrees
of freedom are treated as classical variables. This implies that for
the bonds between atoms in the host we use the Hamiltonian
(\ref{eq:Hhost}) and neglect quantum fluctuations, i.e., in the spin
sector we keep only the $z$th (Ising) spin components and in the orbital 
one only the terms which are proportional to the doublon occupation
numbers (\ref{eq:dub}) and to the identity operators. Similarly, for
the impurity-host bonds we use the Hamiltonian Eq. (\ref{eq:H123})
by keeping only the $z$th projections of spin operators. Since we do
neglect the fluctuation of the spin amplitude it is enough to consider
only the maximal and minimal values of $\langle S^z_i\rangle$ for spin
$S=3/2$ at the impurity sites and $S=1$ at the host atoms. With these
assumptions we can construct all the possible configurations by varying
the spin and orbital configurations at the sites from $i=1$ to $i=4$ in
the cluster shown in Fig. \ref{fig:setup}. Note that the outer ions in 
the cluster belong all to the same sublattice, so two distinct cases 
have to be considered to probe all the configurations. Since physically 
it is unlikely that a single impurity will change the orbital order of
the host globally thus we will not compare the energies from these two
cases and analyze two classes of solutions separately,
see Sec. \ref{sec:two}. Then, the lowest energy configuration in each
class provides the optimal spin-orbital pattern for the NNs around
the $3d$ impurity. In the case of degenerate classical states, the
spin-orbital order is established by including quantum fluctuations.

In the case of a periodic doping analyzed in Sec. \ref{sec:dop}, 
we use a similar strategy in the computation. Taking the most general 
formulation, we employ larger clusters having both size and shape that 
depend on the impurity distribution and on the spin-orbital order in 
the host. For this purpose, the most natural choice is to search for 
the minimum energy configuration in the elementary unit cell that can 
reproduce the full lattice by a suitable choice of the translation 
vectors. This is computationally expensive but doable for a periodic 
distribution of the impurities that is commensurate to the lattice 
because it yields a unit cell of relatively small size for doping around 
$x=0.1$. Otherwise, for the incommensurate doping the size of the unit 
cell can lead to a configuration space of a dimension that impedes
finding of the ground state. This problem is computationally more
demanding and to avoid the comparison of all the energy configurations,
we have employed the Metropolis algorithm at low temperature to achieve
the optimal configuration iteratively along the convergence process.
Note that this approach is fully classical, meaning that the
spins of the host and impurity are treated as Ising variables and the
orbital fluctuations in the host's Hamiltonian Eq. (\ref{eq:Hhost})
are omitted. They will be addressed in Sec. \ref{sec:orb}.

\subsection{Two nonequivalent $3d$ doping cases}
\label{sec:two}

The single impurity problem is the key case to start with because it
shows how the short-range spin-orbital correlations are modified around
the $3d$ atom due to the host and host-impurity interactions in Eq. 
(\ref{fullH}). The analysis is performed by fixing the strength between 
Hund's exchange and Coulomb interaction within the host (\ref{Hint}) 
at $\eta_{\rm host}=0.1$, and by allowing for a variation of the ratio
between the host-impurity interaction (\ref{eq:H123}) and the Coulomb
coupling at the impurity site. The choice of $\eta_{\rm host}=0.1$ is
made here because this value is within the physically relevant range
for the case of the ruthenium materials. Small variations of
$\eta_{\rm host}$ do not affect the obtained results qualitatively.

As we have already discussed in the model derivation, the sign of the
magnetic exchange between the impurity and the host depends on the
orbital occupation of the $4d$ doublon around the $3d$ impurity. The
main aspect that controls the resulting magnetic configuration is then
given by the character of the doublon orbitals around the impurity,
depending on whether they are active or inactive along the considered
$3d-4d$ bond. To explore such a competition quantitatively we
investigate $G$-AO order for the host with alternation of $a$ and $c$
doublon configurations accompanied by the $C$-AF spin pattern, see Fig.
\ref{fig:host}. Note that the $a$ orbitals are active only along the
$b$ axis, while the $c$ orbitals are active along the both axes: $a$
and $b$ \cite{Dag08}. This state has the lowest energy for the host in
a wide range of parameters for Hund's exchange, Coulomb element and
crystal-field potential \cite{Cuo06}.

Due to the specific orbital pattern of Fig. \ref{fig:host}, the $3d$
impurity can substitute one of two distinct $4d$ sites which are
considered separately below, either with $a$ or with $c$ orbital
occupied by the doublon. Since the two $4d$ atoms have
nonequivalent surrounding orbitals, not always active along the $3d-4d$
bond, we expect that the resulting ground state will have a modified
spin-orbital order. Indeed, if the $3d$ atom replaces the $4d$ one with
the doublon in the $a$ orbital, then all the $4d$ neighboring sites
have active doublons along the connecting $3d-4d$ bonds because they
are in the $c$ orbitals. On the contrary,
the substitution at the $4d$ site with $c$ orbital doublon
configuration leads to an impurity state with its neighbors having both
active and inactive doublons. Therefore, we do expect a more intricate
competition for the latter case of an impurity occupying the $4d$ site
with $c$ orbital configuration. Indeed, this leads to frustrated
host-impurity interactions, as we show in Sec. \ref{sec:impc}.

\subsection{Doping removing a doublon in $a$ orbital}
\label{sec:impa}

\begin{figure}[t!]
\includegraphics[width=1\columnwidth]{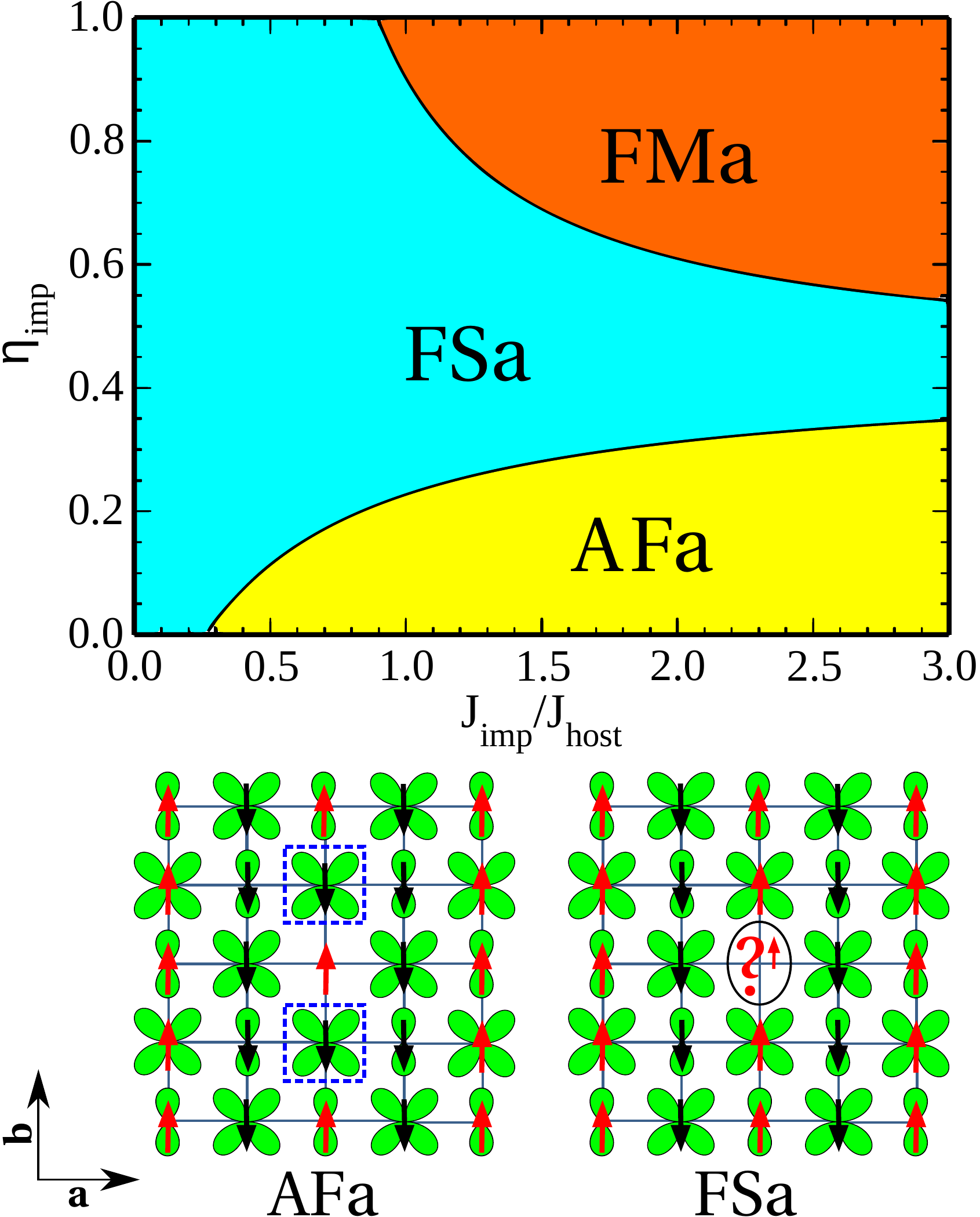}
\caption{
Top --- Phase diagram of the $3d$ impurity in the $(a,b)$ plane with 
the $C$-AF/$G$-AO order in $4d$ host for the impurity doped at the 
sublattice with an $a$-orbital doublon. Different colors refer to local
spin order around the impurity, AF and FM, while FS indicates the 
intermediate regime of frustrated impurity spin.
Bottom --- Schematic view of spin-orbital patterns for the ground state
configurations shown in the top panel. The $3d$ atom is at the central 
site, the dotted frame highlights the $4d$ sites where the impurity 
induces a a spin reversal. In the FS$a$ phase the question mark stands
for that the frustrated impurity spin within the classical approach but
frustration is released by the quantum fluctuations of the NN $c$
orbitals in the $a$ direction resulting in small AF couplings along the
$a$ axis, and spins obey the $C$-AF order (small arrow). The labels
FM$a$ and AF$a$ refer to the local spin order around the
$3d$ impurity site with respect to the host
--- these states differ by spin inversion at the $3d$ atom site.
\label{fig:pd_caf_a}}
\end{figure}

We start by considering the physical situation where the $3d$ impurity
replaces a $4d$ ion with the doublon within the $a$ orbital. The ground 
state phase diagram and the schematic view of the spin-orbital
pattern are reported in Fig. \ref{fig:pd_caf_a} in terms of the ratio
$J_{\rm imp}/J_{\rm host}$ (\ref{eq:etai}) and the strength of Hund's 
exchange coupling $\eta_{\rm imp}$ (\ref{eq:bothJ}) at the $3d$ site.
There are three different ground states that appear in the phase
diagram. Taking into account the structure of the $3d-4d$ spin-orbital
exchange (\ref{eq:H123}) we expect that, in the regime where the
host-impurity interaction is greater than that in the host, the $4d$
neighbors to the impurity tend to favor the spin-orbital configuration
set by the $3d-4d$ exchange. In this case, since the orbitals
surrounding the $3d$ site already minimize the $3d-4d$ Hamiltonian, we
expect that the optimal spin configuration corresponds to the $4d$
spins aligned either antiferromagnetically or ferromagnetically
with respect to the impurity $3d$ spin.

The neighbor spins are AF to the $3d$ spin impurity in the AF$a$ phase,
while the FM$a$ phase is just obtained from AF$a$ by reversing the spin
at the impurity, and having all the $3d-4d$ bonds FM. It is interesting
to note that due to the host-impurity interaction the $C$-AF spin
pattern of the host is modified in both the AF$a$ and the FM$a$ ground
states. Another intermediate configuration which emerges when the
host-impurity exchange is weak in the intermediate FS$a$ phase where
the impurity spin is undetermined and its configuration in the initial
$C$-AF phase is degenerate with the one obtained after
the spin-inversion operation. This is a singular physical situation
because the impurity does not select a specific direction even if the
surrounding host has a given spin-orbital configuration. Such a
degeneracy is clearly verified at the critical point
$\eta_{\rm imp}^c\simeq 0.43$ where the amplitude of the $3d-4d$
coupling vanishes when the doublon occupies the active orbital.
Interestingly, such a degenerate
configuration is also obtained at $J_{\rm imp}/J_{\rm host}<1$ when the
host dominates and the spin configuration at the $4d$ sites around the
impurity are basically determined by $J_{\rm host}$. In this case, due
to the $C$-AF spin order, always two bonds are FM and other two have AF
order, independently of the spin orientation
at the $3d$ impurity. This implies that both FM or AF couplings along
the $3d-4d$ bonds perfectly balance each other which results in the
degenerate FS$a$ phase.

\begin{figure}[t!]
\includegraphics[width=0.9\columnwidth]{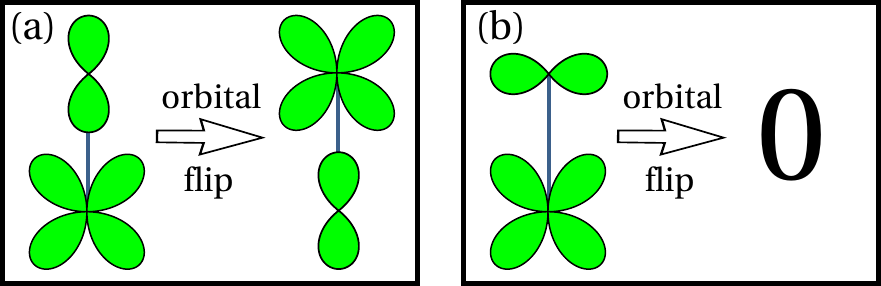}
\caption{
Schematic view of the two types of orbital bonds found in the $4d$ host:
(a) an active bond with respect to orbital flips,
$(\tau_{i}^{\gamma +}\tau_{j}^{\gamma -}+{\rm H.c.})$, and
(b) an inactive bond, where orbital fluctuations are blocked by the
orbital symmetry --- here the orbitals are static and
only Ising terms contribute to the ground state energy.}
\label{fig:host_flips}
\end{figure}

It is worth pointing out that there is a quite large region of the 
phase diagram where the FS$a$ state is stabilized and the spin-orbital 
order of the host is not affected by doping with the possibility of 
having large degeneracy in the spin configuration of the impurities. 
On the other hand, by inspecting the $c$ orbitals around the impurity 
(Fig. \ref{fig:pd_caf_a}) from the point of view of the full host's 
Hamiltonian Eq. (\ref{eq:Hhost}) with orbital flips included,
$(\tau_{i}^{\gamma +}\tau_{j}^{\gamma -}+{\rm H.c.})$, one can easily
find out that the frustration of the impurity spin can be released by
quantum orbital fluctuations. Note that the $c$ orbitals around the
impurity in the $a$ ($b$) direction have quite different surroundings. 
The ones along the $a$ axis are connected by two active bonds along 
the $b$ axis with orbitals $a$, as in Fig. \ref{fig:host_flips}(a),
while the ones along $b$ are connected with \textit{only one} active 
$a$ orbital along the same $b$ axis. This means that in the perturbative 
expansion the orbital flips will contribute only along the $b$ bonds
(for the present $G$-AO order) and admix the $a$ orbital character to 
$c$ orbitals along them, while such processes will be blocked for the 
bonds along the $a$ axis, as also for $b$ orbitals along the $b$ axis, 
see Fig. \ref{fig:host_flips}(b). 

This fundamental difference can be easily included in the host-impurity
bond in the mean-field manner by setting
$\langle D_{i\pm b,\gamma}\rangle=0$ for the bonds along the $b$ axis
and $0<\langle D_{i\pm a,\gamma}\rangle<1$ for the bonds along the $a$
axis. Then one can easily check that for the impurity spin pointing
downwards we get the energy contribution from the spin-spin bonds which
is given by
$E_\downarrow=\alpha(\eta_{\rm host})\langle D_{i\pm a,\gamma}\rangle$,
and for the impurity spin pointing upwards we have
$E_\uparrow=-\alpha(\eta_{\rm host})\langle D_{i\pm a,\gamma}\rangle$,
with $\alpha(\eta_{\rm host})>0$. Thus, it is clear that any admixture
of the virtual orbital flips in the host's wave function polarize the
impurity spin upwards so that the $C$-AF order of the host will be
restored.

\subsection{Doping removing a doublon in $c$ orbital}
\label{sec:impc}

Let us move to the case of the $3d$ atom replacing the doublon at $c$
orbital. As anticipated above, this
configuration is more intricate because the orbitals surrounding the
impurity, as originated by the $C$-AF/AO order within the host, lead
to nonequivalent $3d-4d$ bonds. There are two bonds with the doublon
occupying an inactive orbital (and has no hybridization with the
$t_{2g}$ orbitals at $3$d atom), and two remaining bonds with doublons
in active $t_{2g}$ orbitals.

\begin{figure}[t!]
\includegraphics[width=1\columnwidth]{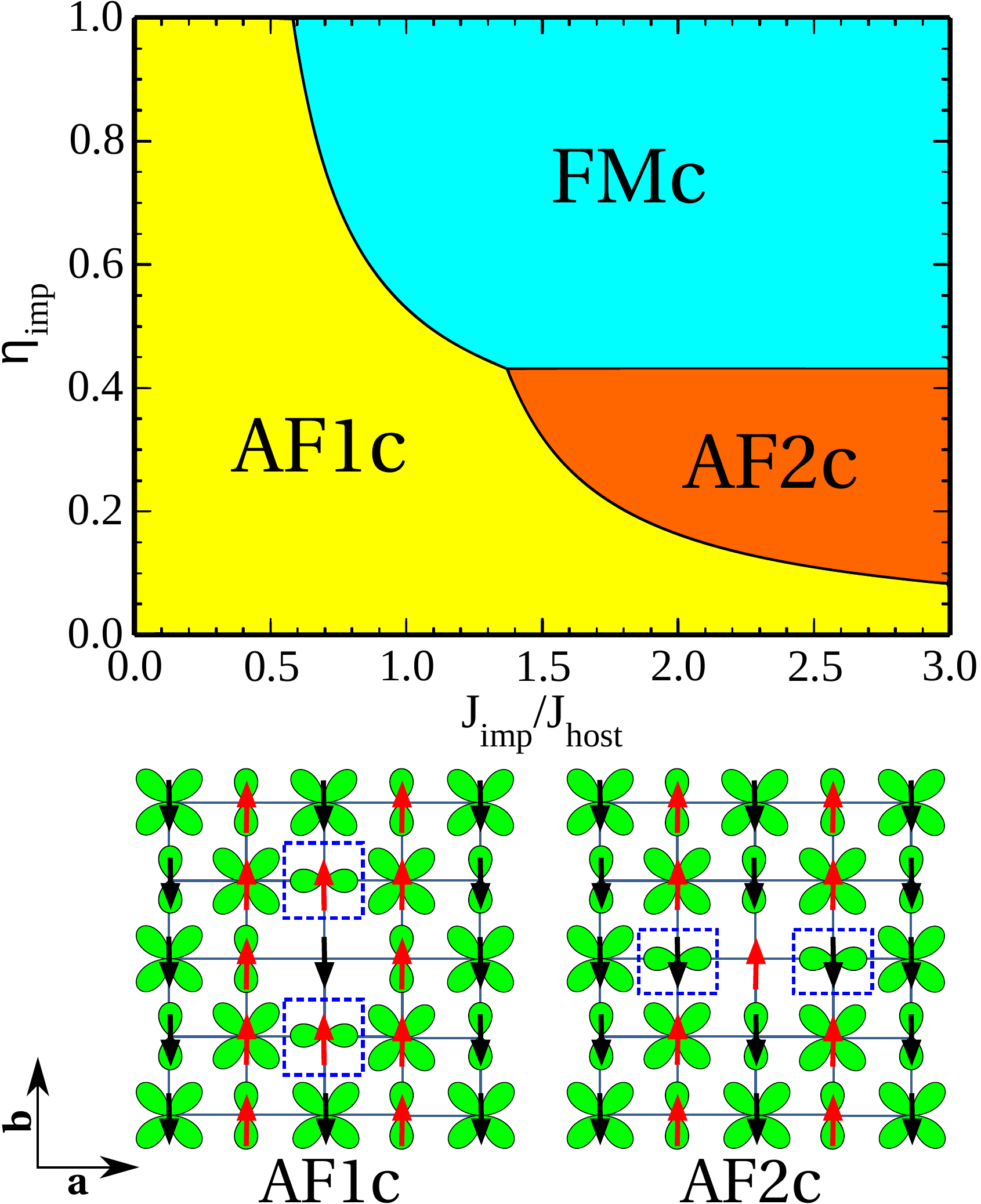}
\caption{
Top --- Phase diagram of the $3d$ impurity in the $4d$ host with
$C$-AF/$G$-AO order and the impurity doped at the $c$ doublon
sublattice. Different colors refer to local spin order around
the impurity: AF$c1$, AF$c2$, and FM.
Bottom --- Schematic view of spin-orbital patterns for the two AF 
ground state configurations shown in the top panel; the FM$c$ phase 
differs from the AF$c2$ one only by spin inversion at the $3d$ atom.
The $3d$ atom is at the central site and has no doublon orbital, the 
frames highlight the spin-orbital defects caused by the impurity. 
As in Fig. \ref{fig:pd_caf_a}, the labels AF and FM refer to the 
impurity spin orientation with respect to the neighboring $4d$ sites.
}
\label{fig:pd_caf_c}
\end{figure}

Since the $3d-4d$ spin-orbital exchange depends on the orbital
polarization of $4d$ sites we do expect a competition which may modify
significantly the spin-orbital correlations in the host. Indeed, one
observes that three configurations compete, denoted as AF1$c$, AF2$c$
and FM$c$, see Fig. \ref{fig:pd_caf_c}. In the regime where the
host-impurity exchange dominates the system tends to minimize the
energy due to the $3d-4d$ spin-orbital coupling and, thus, the orbitals
become polarized in the active configurations compatible with the
$C$-AF/$G$-AO pattern and the host-impurity spin coupling is AF for
$\eta_{\rm imp}\leq 0.43$, while it is FM otherwise. This region
resembles orbital polarons in doped manganites \cite{Kil99,Dag04}.
Also in this case, the orbital polarons arise because they minimize the
double exchange energy \cite{deG60}.

On the contrary, for weak spin-orbital coupling between the
impurity and the host there is an interesting cooperation between the
$3d$ and $4d$ atoms. Since the strength of the impurity-host coupling
is not sufficient to polarize the orbitals at the $4d$ sites, it is
preferable to have an orbital rearrangement to the configuration with
inactive orbitals on $3d-4d$ bonds and spin flips at $4d$ sites. In
this way the spin-orbital exchange is optimized in the host and also on
the $3d-4d$ bonds. The resulting state has an AF coupling between the
host and the impurity as it should when all the orbitals surrounding
the $3d$ atoms are inactive with respect to the bond direction. This
modification of the orbital configuration induces the change in spin
orientation. The double exchange bonds (with inactive doublon orbitals)
along the $b$ axis are then blocked and the total energy is lowered,
in spite of the frustrated spin-orbital exchange in the host.
As a result, the AF1$c$ state the spins surrounding the impurity are
aligned and antiparallel to the spin at the $3d$ site.

Concerning the host $C$-AF/$G$-AO order we note that it is modified
only along the direction where the FM correlations develop and spin
defects occur within the chain doped by the $3d$ atom. The FM order is
locally disturbed by the $3d$ defect antiferromagnetically coupled
spins surrounding it. Note that this phase is driven by the orbital
vacancy as the host develops more favorable orbital bonds to gain the
energy in the absence of the orbital degrees of freedom at the impurity.
At the same time the impurity-host bonds do not generate too big energy
losses as:
(i) either $\eta_{\rm imp}$ is so small that the loss due to $E_D$ is
compensated by the gain from the superexchange
$\propto J_{S}(D_j^{(\gamma)}=1)$ (all these bonds are AF), see
Fig. \ref{fig:JSEd}, or
(ii) $J_{\rm imp}/J_{\rm host}$ is small meaning that the overall energy
scale of the impurity-host exchange remains small.
Interestingly, if we compare the AF1$c$ with the AF2$c$ ground states we
observe that the disruption of the $C$-AF/$G$-AO order is anisotropic
and occurs either along the FM chains in the AF1$c$ phase or
perpendicular to the FM chains in the AF2$c$ phase. No spin frustration 
is found here, in contrast to the FS$a$ phase in the case of $a$ 
doublon doping, see Fig. \ref{fig:pd_caf_a}.

Finally, we point out that a very similar phase diagram can be obtained
assuming that the host has the FM/$G$-AO order with $a$ and $b$ orbitals
alternating from site to site. Such configuration can be stabilized by a
distortion that favors the out-of-plane orbitals. In this case there is
no difference in doping at one or the other sublattice. The main
difference is found in energy scales --- for the $G$-AF/$C$-AO order the
diagram is similar to the one of Fig. \ref{fig:pd_caf_c} if we rescale
$J_{\rm imp}$ by half, which means that the $G$-AF order is softer than
the $C$-AF one. Note also that in the peculiar AF1$c$ phase the impurity
does not induce any changes in the host for the FM/AO ordered host.
Thus we can safely conclude that the observed change in the orbital
order for the $C$-AF host in the AF1$c$ phase is due to the presence of
the $c$ orbitals which are not directional in the $(a,b)$ plane.

Summarizing, we have shown the complexity of local spin-orbital order
around $t_{2g}^3$ impurities in a $4t_{2g}^4$ host. It is remarkable
that such impurity spins not only modify the spin-orbital order around
them in a broad regime of parameters, but also are frequently
frustrated. This highlights the importance of quantum effects beyond
the present classical approach which release frustration as we show
in Sec. \ref{sec:orb}.

\section{Periodic 3\MakeLowercase{d} doping in 4\MakeLowercase{d} host}
\label{sec:dop}

\subsection{General remarks on finite doping}
\label{sec:dopg}

In this Section we analyze the spin-orbital patterns due to a finite
concentration $x$ of $3d$ impurities within the $4d$ host with
$C$-AF/$G$-AO order, assuming that the $3d$ impurities are distributed 
in a periodic way. The study is performed for three representative 
doping distributions --- the first one $x=1/8$ is commensurate with the 
underlying spin-orbital order and the other two are incommensurate with 
respect to it, meaning that in such cases doping at both $a$ and $c$ 
doublon sites is imposed simultaneously.

As the impurities lead to local energy gains due to $3d-4d$ bonds 
surrounding them, we expect that the most favorable situation is when 
they are isolated and have maximal distances between one another. 
Therefore, we selected the largest possible distances for the three 
doping levels used in our study: $x=1/8$, $x=1/5$, and $x=1/9$. This 
choice allows us to cover different regimes of competition between the 
spin-orbital coupling within the host and the $3d-4d$ coupling. While 
single impurities may only change spin-orbital order locally, we use 
here a high enough doping to investigate possible global changes in
spin-orbital order, i.e., whether they can occur in the respective 
parameter regime. The analysis is performed as for a single impurity, 
by assuming the classical spin and orbital variables and by determining 
the configuration with the lowest energy.
For this analysis we set the spatial distribution of the $3d$ atoms and
we determine the spin and orbital profile that minimizes the energy.

\subsection{$C$-AF phase with $x=1/8$ doping}
\label{sec:dop8}

\begin{figure}[t!]
\includegraphics[clip,width=1\columnwidth]{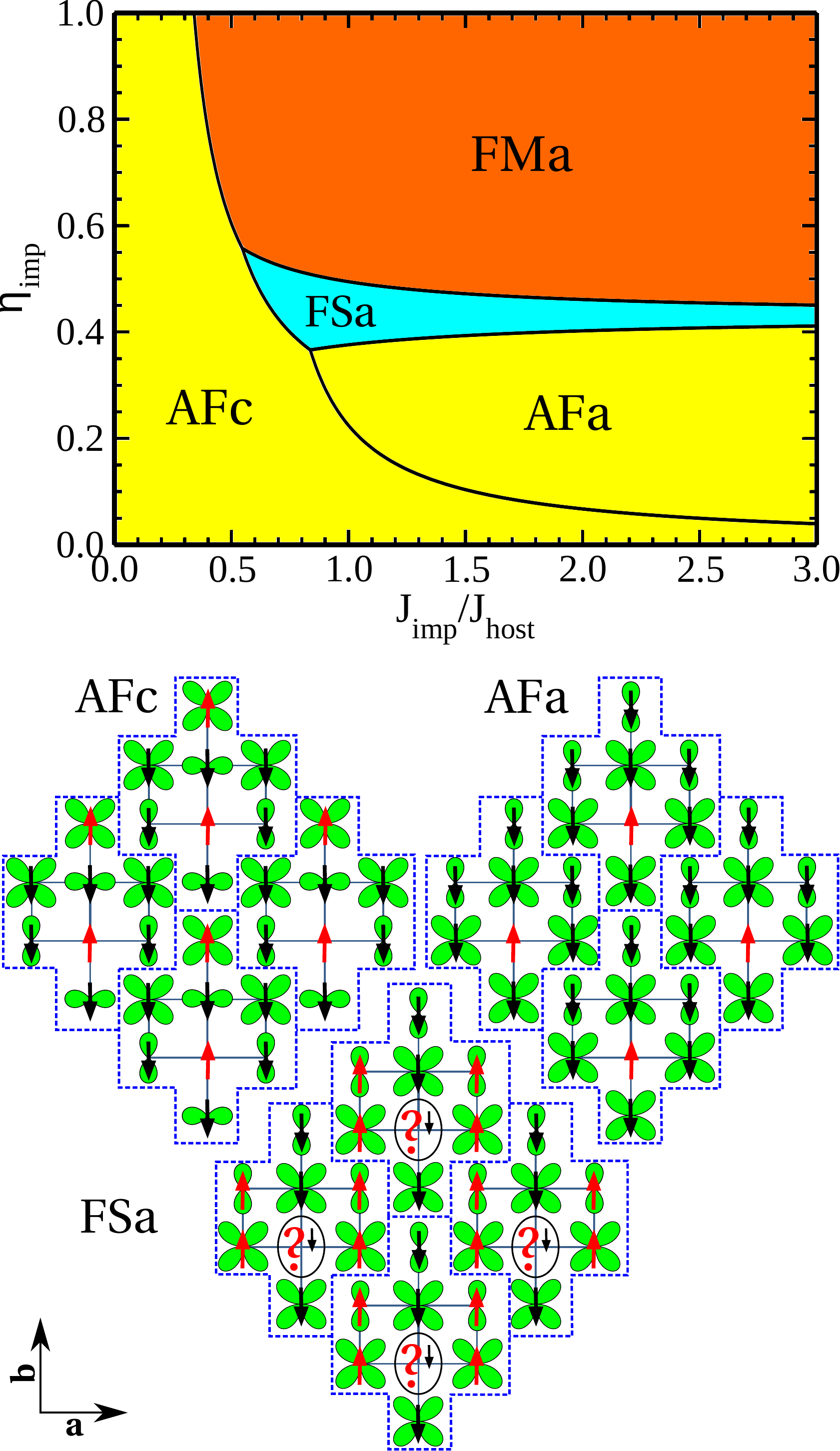}
\caption{
Top panel --- Ground state diagram obtained for periodic $3d$ doping
$x=1/8$. Different colors refer to local spin order around
the impurity: AF$a$, AF$c$, FS$a$, and FM$a$.
Bottom panel --- Schematic view of the ground state configurations
within the four 8-site unit cells (indicated by blue dashed lines)
for the phases shown in the phase diagram. The question marks in FS$a$ 
phase indicate frustrated impurity spins within the classical approach 
--- the spin direction (small arrows) is fixed only by quantum 
fluctuations.
The $3d$ atoms are placed at the sites where orbitals are absent.}
\label{fig:1to8}
\end{figure}

We begin with the phase diagram obtained at $x=1/8$ $3d$ doping, see
Fig. \ref{fig:1to8}. In the regime of strong impurity-host coupling the
$3d-4d$ spin-orbital exchange determines the orbital and spin
configuration of the $4d$ atoms around the impurity. The most favorable
state is when the doublon occupies $c$ orbitals at the NN sites to the
impurity. The spin correlations between the impurity and the host are
AF (FM), if the amplitude of $\eta_{\rm imp}$ is below (above)
$\eta_{\rm imp}^c$, leading to the AF$a$ and the FM$a$ states, see Fig.
\ref{fig:1to8}. The AF$a$ state has a striped-like profile with AF 
chains alternated by FM domains (consisting of three chains) along the 
diagonal of the square lattice. Even if the coupling between the 
impurity and the host is AF for all the bonds in the AF$a$ state, the 
overall configuration has a residual magnetic moment originating by the 
uncompensated spins and by the cooperation between the spin-orbital 
exchange in the $4d$ host and that for the $3d-4d$ bonds. Interestingly, 
at the point where the dominant $3d-4d$ exchange tends to zero
(i.e., for $\eta_{\rm imp}\simeq\eta_{\rm imp}^c$),
one finds a region of the FS$a$ phase which is analogous to the FS$a$
phase found in Sec. \ref{sec:impa} for a single impurity, see
Fig. \ref{fig:pd_caf_a}. Again the impurity spin is frustrated in
purely classical approach but this frustration is easily released
by the orbital fluctuations in the host so that the $C$-AF order of
the host can be restored. This state is stable for the amplitude of
$\eta_{\rm imp}$ being close to $\eta_{\rm imp}^c$.

The regime of small $J_{\rm imp}/J_{\rm host}$ ratio is qualitatively
different --- an orbital rearrangement around the impurity takes place,
with a preference to move the doublons into the inactive orbitals along
the $3d-4d$ bonds. Such orbital configurations favor the AF spin
coupling at all the $3d-4d$ bonds which is stabilized by the $4d-4d$
superexchange \cite{Fei99}. This configuration is peculiar because it
generally breaks inversion and does not have any plane of mirror
symmetry. It is worth pointing out that the original order in the $4d$
host is completely modified by the small concentration of $3d$ ions
and one finds that the AF coupling between the $3d$ impurity and
the $4d$ host generally leads to patterns such as the AF$c$ phase where
FM chains alternate with AF ones in the $(a,b)$ plane. Another relevant
issue is that the cooperation between the host and impurity can lead to
a fully polarized FM$a$ state. This implies that doping can release the
orbital frustration which was present in the host with the $C$-AF/$G$-AO
order.

\subsection{Phase diagram for periodic $x=1/5$ doping}
\label{sec:dop5}

\begin{figure*}[t!]
\includegraphics[clip,width=0.90\textwidth]{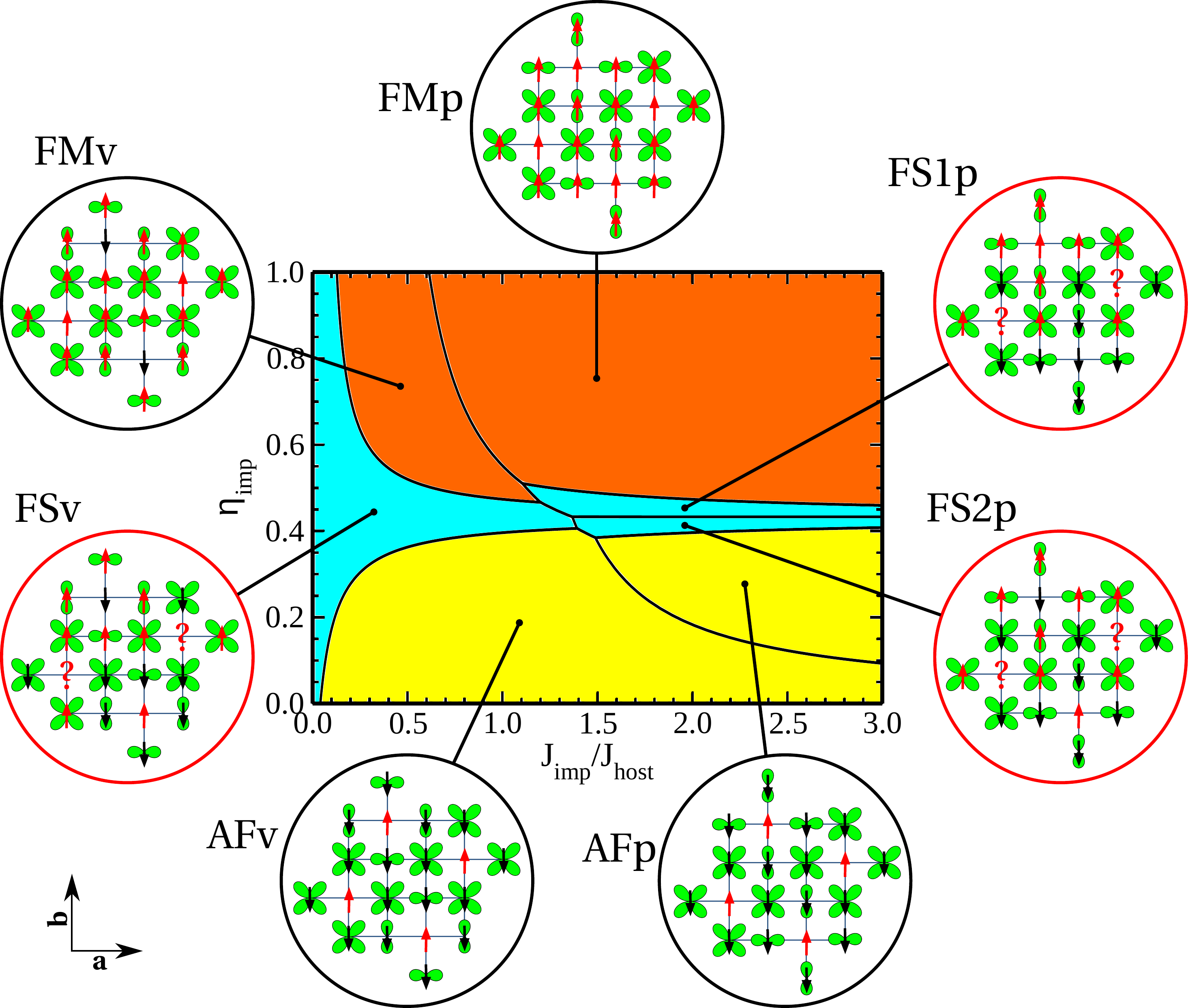}
\caption{ Ground state diagram for $x=1/5$ periodic
concentration of $3d$ impurities (sites where orbitals are absent) with
schematic views of the ground state configurations obtained for the
unit cell of $20$ sites. Spin and orbital order are shown by arrows and
orbitals occupied by doublons; magnetic phases (AF, FS, and FM) are
highlighted by different color. The question marks in FS states
(red circles) indicate frustrated impurity spins within the classical
approach.
\label{fig:1to5}}
\end{figure*}

\begin{figure}[b!]
\includegraphics[clip,width=.96\columnwidth]{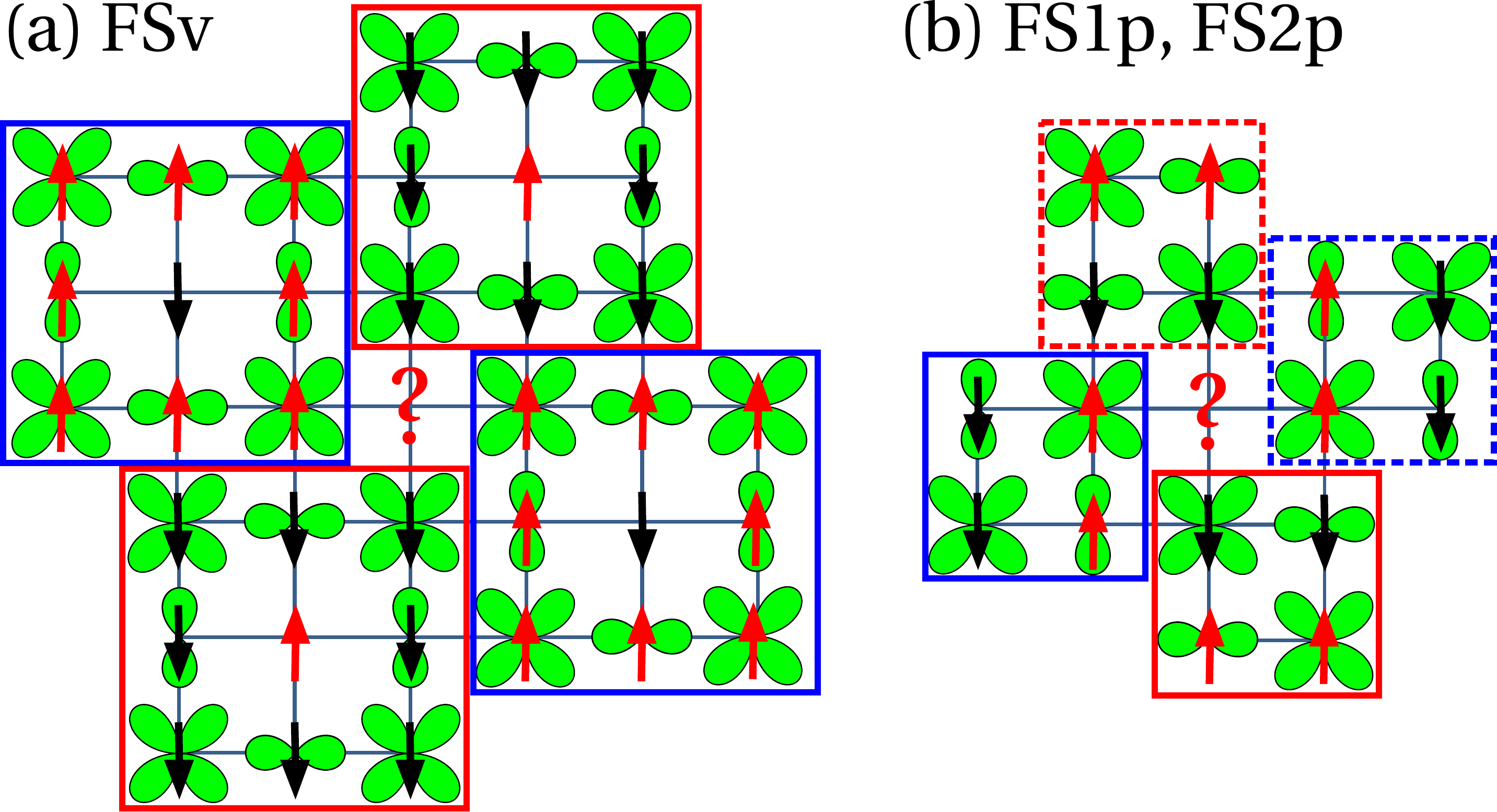}
\caption{ Isotropic surrounding of the degenerate impurity spins in the 
FS$v$ and FS$p$ phases in the case of $x=1/5$ periodic doping (Fig.
\ref{fig:1to5}). Frames mark the clusters which are not connected with
orbitally active bonds.
\label{fig:FSv}}
\end{figure}

Next we consider doping $x=1/5$ with a given periodic spatial profile
which concerns both doublon sublattices. We investigate the $3d$ spin 
impurities separated by the translation vectors $\vec{u}=(i,j)$ and 
$\vec{v}=(2,-1)$ (one can show that for general periodic doping $x$, 
$|\vec{u}|^2=x^{-1}$) so there is a mismatch between the impurity 
periodicity and the two-sublattice $G$-AO order in the host. One finds 
that the present case, see Fig. \ref{fig:1to5}, has similar general 
structure of the phase diagram to the case of $x=1/8$ (Fig. 
\ref{fig:1to8}), with AF correlations dominating for $\eta_{\rm imp}$ 
lower than $\eta_{\rm imp}^c$ and FM ones otherwise. Due to the specific 
doping distribution there are more phases appearing in the ground state
phase diagram. For $\eta_{\rm imp}<\eta_{\rm imp}^c$ the most stable
spin configuration is with the impurity coupled antiferromagnetically
to the host. This happens both in the AF vacancy (AF$v$) and the AF 
polaronic (AF$p$) ground states. The difference between the two AF 
states arises due to the orbital arrangement around the impurity.
For weak ratio of the impurity to the host spin-orbital exchange,
$J_{\rm imp}/J_{\rm host}$, the orbitals around the impurity are all
inactive ones. On the contrary, in the strong impurity-host coupling 
regime all the orbitals are polarized to be in active (polaronic) 
states around the impurity. Both states have been found as AF1$c$ and 
AF2$c$ phase in the single impurity problem (Fig. \ref{fig:pd_caf_c}).

More generally, for all phases the boundary given by an approximate
hyperbolic relation $\eta_{\rm imp}\propto J_{\rm imp}^{-1}$ separates
the phases where the orbitals around impurities in the $c$-orbital
sublattice are all inactive (small $\eta_{\rm imp}$) from those where
all the orbitals are active (large $\eta_{\rm imp}$). The inactive
orbital around the impurity stabilize always the AF coupling between
the impurity spin and host spins whereas the active orbitals can give
either AF or FM exchange depending on $\eta_{\rm imp}$ (hence
$\eta_{\rm imp}^c$, see Fig. \ref{fig:JSEd}). Since the doping does
not match the size of the elementary unit cell, the resulting ground
states do not exhibit specific symmetries in the spin-orbital pattern.
They are generally FM due to the uncompensated magnetic moments and
the impurity feels screening by the presence of the surrounding it host
spins being antiparallel to the impurity spin.

\begin{figure*}[t!]
\includegraphics[clip,width=0.92\textwidth]{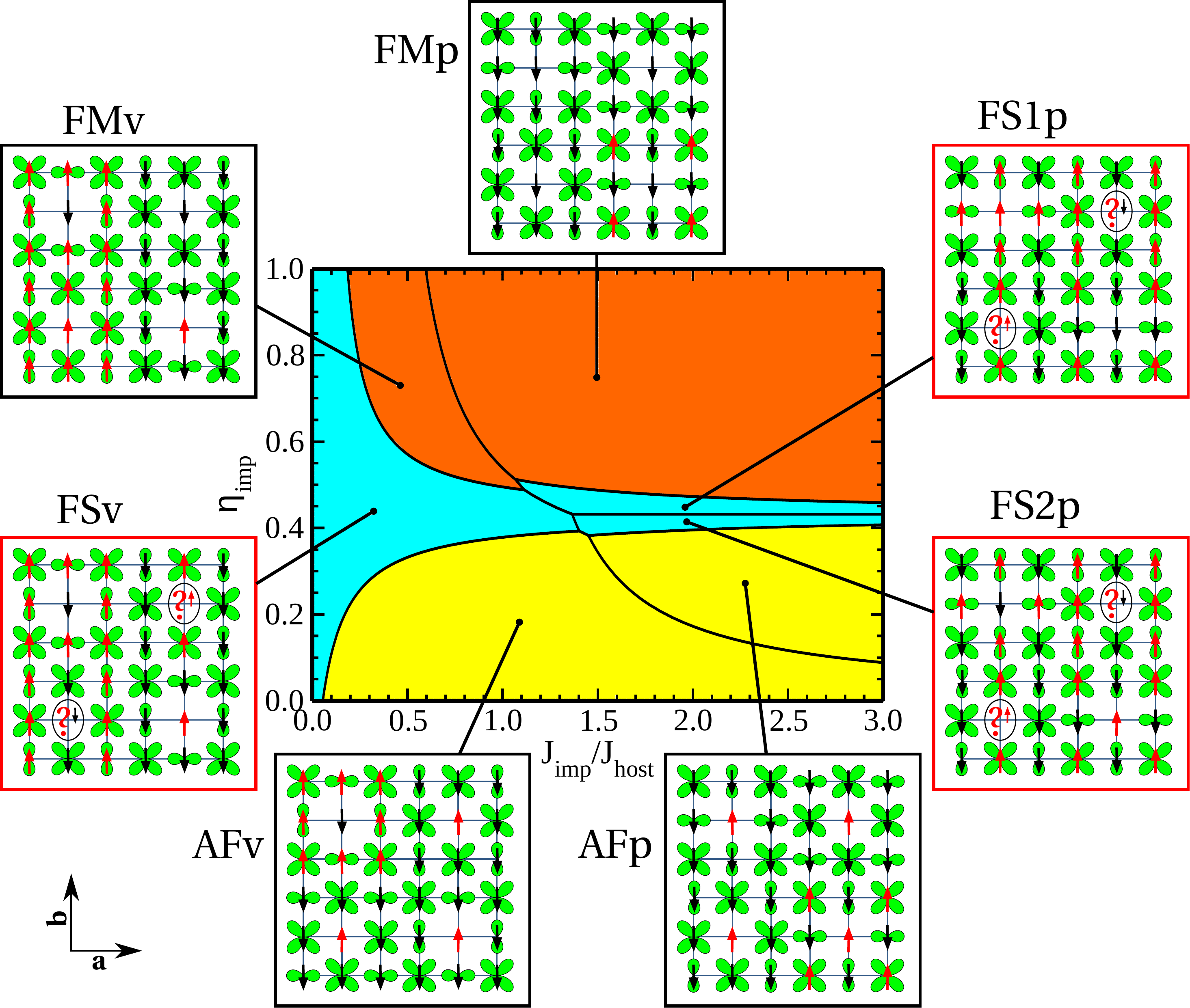}
\caption{ Ground state diagram for $x=1/9$ periodic
concentration of $3d$ impurities with schematic views of the ground
state configurations obtained for the unit cell of $36$ sites.
Spin order (AF, FS, and FM) is highlighted by different color.
The question marks in FS states (red squares) indicate frustrated
impurity spins within the classical approach ---
the spin is fixed here by quantum fluctuations (small arrows).
Doped $3d$ atoms are at the sites where orbitals are absent.}
\label{fig:1to9}
\end{figure*}

By increasing Hund's exchange coupling at the $3d$ ion the system
develops fully FM state in a large region of the ground state diagram
due to the possibility of suitable orbital polarization around the
impurity. On the other hand, in the limit where the impurity-host bonds
are weak, so either for $\eta_{\rm imp}\simeq\eta_{\rm imp}^c$ and
large enough $J_{\rm imp}/J_{\rm host}$ so that all orbitals around the
impurity are active, or just for small $J_{\rm imp}/J_{\rm host}$ we
get the FS phases where the impurity spin at the $a$-orbital sublattice
is undetermined in the present classical approach. This is a similar 
situation to the one found in the FS$a$ phase of a single impurity 
problem and at $x=1/8$ periodic doping, see Figs. \ref{fig:pd_caf_a} 
and \ref{fig:1to8}, but there it was still possible to identify the 
favored impurity spin polarization by considering the orbital flips 
in the host around the impurity.

However, the situation here is different as the host's order is
completely altered by doping and has became isotropic, in contrast to
the initial $C$-AF order (Fig. \ref{fig:host}) which breaks the 
planar symmetry between the $a$ and $b$ direction. It was precisely this
symmetry breaking that favored one impurity spin polarization over the
other one. Here this mechanism is absent --- one can easily check that
the neighborhood of the $c$ orbitals surrounding impurities is
completely equivalent in both directions (see Fig. \ref{fig:FSv} for
the view of these surroundings) so that the orbital flip argument is no
longer applicable. This is a peculiar situation in the classical
approach and we indicate frustration in spin direction by question
marks in Fig. \ref{fig:1to5}.

In Fig. \ref{fig:FSv} we can see that both in FS vacancy (FS$v$) and 
FS polaronic (FS$p$) phase the orbitals are grouped in $3\times 3$ 
clusters and $2\times 2$ plaquettes, respectively, that encircle the 
degenerate impurity spins. For the FS$v$ phase we can distinguish 
between two kind of plaquettes with non-zero spin polarization 
differing by a global spin inversion. In the case of FS$p$ phases we 
observe four plaquettes with zero spin polarization arranged in two 
pairs related by a point reflection with respect to the impurity site. 
It is worthwhile to realize that these plaquettes are completely 
disconnected in the orbital sector, i.e., there are no orbitally 
active bonds connecting them (see Fig. \ref{fig:host_flips} for the 
pictorial definition of orbitally active bonds). 
This means that quantum effects of purely orbital nature can appear 
only at the short range, i.e., inside the plaquettes. However, one 
can expect that if for some reason the two degenerate spins in a
single elementary cell will polarize then they will also polarize in 
the same way in all the other cells to favor long-range quantum 
fluctuations in the spin sector related to the translational 
invariance of the system.

\subsection{Phase diagram for periodic $x=1/9$ doping}
\label{sec:dop9}

Finally we investigate low doping $x=1/9$ with a given periodic spatial 
profile, see Fig. \ref{fig:1to9}. Here the impurities are separated by 
the translation vectors $\vec{u}=(0,3)$ and $\vec{v}=(3,0)$. Once again
there is a mismatch between the periodic distribution of impurities and
the host's two-sublattice AO order, so we again call this doping 
incommensurate as it also imposes doping at both doublon sublattices. 
The ground state diagram presents gradually increasing tendency towards 
FM $3d-4d$ bonds with increasing $\eta_{\rm imp}$, see Fig. 
\ref{fig:1to9}. These polaronic bonds polarize as well the $4d-4d$ 
bonds and one finds an almost FM order in the FM$p$ state. Altogether, 
we have found the same phases as at the higher doping of $x=1/5$, see 
Fig. \ref{fig:1to5}, i.e., AF$v$ and AF$p$ at low values $\eta_{\rm imp}$,
FM$v$ and FM$p$ in the regime of high $\eta_{\rm imp}$, separated by the 
regime of frustrated impurity spins which occur within the phases:
FS$v$, FS1$p$, and FS2$p$.

The difference between the two AF (FM) states in Fig. \ref{fig:1to9} is
due to the orbital arrangement around the impurity. As for the other 
doping levels considered so far, $x=1/8$ and $x=1/5$, we find neutral
(inactive) orbitals around $3d$ impurities in the regime of low
$J_{\rm imp}/J_{\rm host}$ in AF$v$ and FM$v$ phases which lead to spin
defects within the 1D FM chains in the $C$-AF spin order.
A similar behavior was reported for single impurities in the low doping
regime in Sec. \ref{sec:imp}. This changes radically above the orbital
transition for both types of local magnetic order, where the orbitals
reorient into the active ones. One finds that spin orientations are
then the same as those of their neighboring $4d$ atoms, with some
similarities to those found at $x=1/5$, see Fig. \ref{fig:1to5}.

Frustrated impurity spins occur in the crossover regime between the AF
and FM local order around impurities. This follows from the local
configurations around them, which include two $\uparrow$-spins and two
$\downarrow$-spins accompanied by $c$ orbitals at the NN $4d$ sites.
This frustration is easily removed by quantum fluctuations and we
suggest that this happens again in the same way as for $x=1/8$ doping,
as indicated by small arrows in the respective FS phases shown in Fig.
\ref{fig:1to9}.

\section{Quantum effects beyond the classical approach}
\label{sec:qua}

\subsection{Spin-orbital quantum fluctuations}
\label{sec:orb}

\begin{figure*}[t!]
\begin{centering}
\includegraphics[width=.99\textwidth]{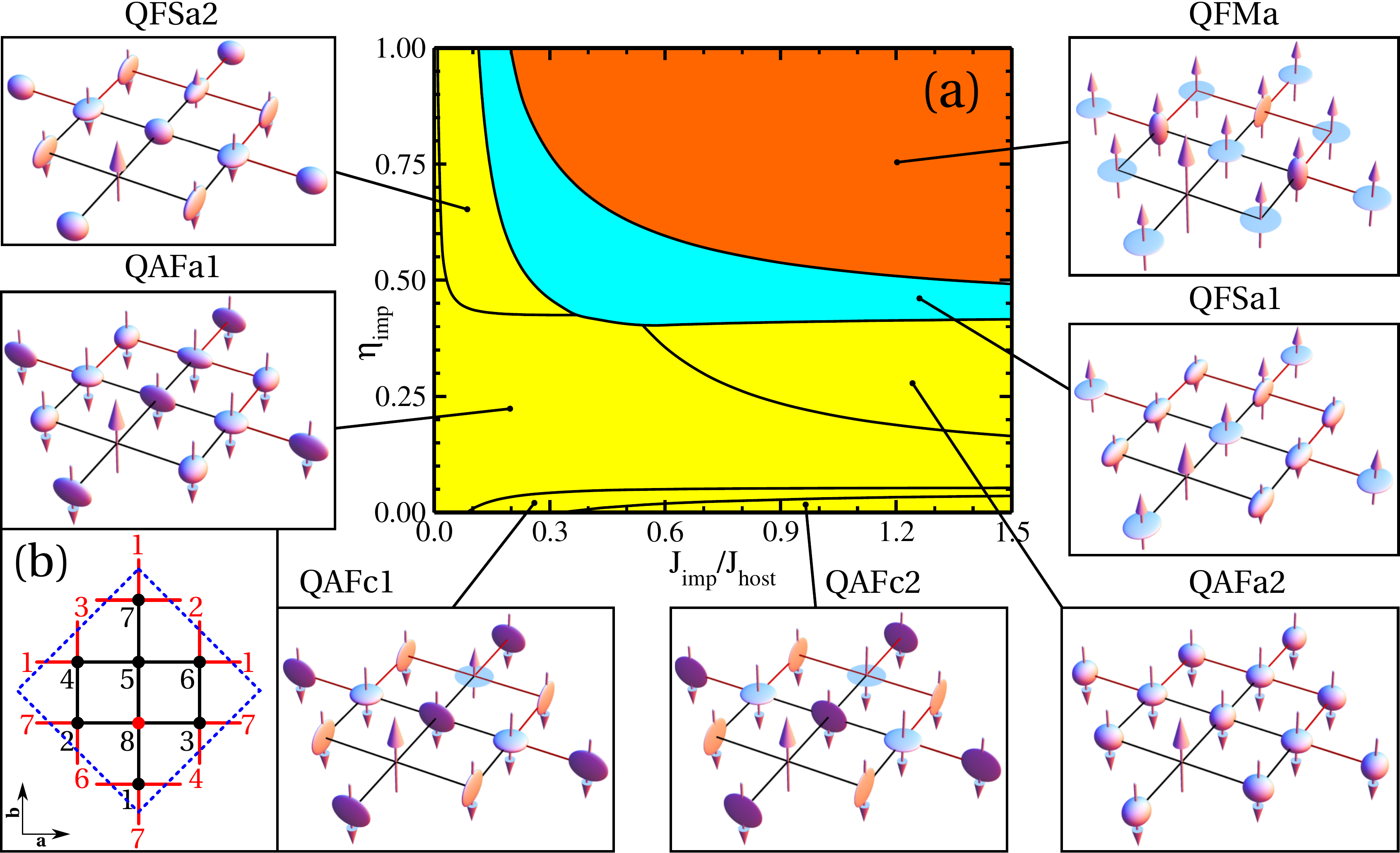}
\par\end{centering}
\protect\caption{
(a) Phase diagram for the quantum problem at zero spin-orbit simulated
on the 8-site cluster in the presence of one-impurity. Arrows and
ellipsoids indicate the spin-orbital state at a given site $i$, while
the shapes of ellipsoids reflect the orbital avarages:
$\langle a^{\dag}_ia_i^{}\rangle$, $\langle b^{\dag}_ib_i^{}\rangle$
and $\langle c^{\dag}_ic_i^{}\rangle$ (i.e., a circle in the plane
perpendicular to the axis $\gamma$ implies 100\% occupation of the
orbital $\gamma$).
(b) The periodic cluster of $L=8$ sites used, with the orbital
dilution ($3d^3$ impurity) at site $i=8$. The dotted lines identify 
the basic unit cell adopted for the simulation with the same symmetries 
of the square lattice. }
\label{fig:diag}
\end{figure*}

So far, we analyzed the ground states of $3d$ impurities in the $(a,b)$
plane of a $4d$ system using the classical approach. Here we show that
this classical picture may be used as a guideline and is only
quantitatively changed by quantum fluctuations if the spin-orbit 
coupling is weak. We start the analysis by
considering the quantum problem in the absence of spin-orbit coupling
(at $\lambda=0$). The orbital doublon densities,
\begin{equation}
N_{\gamma}\equiv\sum_{i\in\rm host}\langle n_{i\gamma}\rangle,
\label{doc}
\end{equation}
with $\gamma=a,b,c$, and total $S^z$ are conserved quantities and thus
good quantum numbers for a numerical simulation. To determine the
ground state configurations in the parameters space and the relevant
correlation functions we diagonalize exactly the Hamiltonian matrix  
(\ref{fullH}) for the cluster of $L=8$ sites by means of the Lanczos
algorithm. In Fig. \ref{fig:diag}(a) we
report the resulting quantum phase diagram for an 8-site cluster having
one impurity and assuming periodic boundary conditions, see Fig.
\ref{fig:diag}(b). This appears to be an optimal cluster configuration
because it contains a number of sites and connectivities that allows us
to analyze separately the interplay between the host-host and the
host-impurity interactions and to simulate a physical situation when
the interactions within the host dominate over those between the host
and the impurity. Such a problem is a quantum analogue of the single
unit cell presented in Fig. \ref{fig:1to8} for $x=1/8$ periodic doping.

As a general feature that resembles the classical phase diagram, we
observe that there is a prevalent tendency to have AF-like (FM-like)
spin correlations between the impurity and the host sites in the region
of $\eta_{\rm imp}$ below (above) the critical point at
$\eta_{\rm imp}^c\simeq 0.43$ which separates these two regimes, with
intermediate configurations having frustrated magnetic exchange.
As we shall discuss below it is the orbital degree of freedom that
turns out to be more affected by the quantum effects. Following the
notation used for the classical case, we distinguish various
quantum AF (QAF) ground states, i.e., QAF$cn$ ($n=1,2$) and QAF$an$
($n=1,2$), as well as a uniform quantum FM (QFM) configuration, i.e.,
QFM$a$, and quantum frustrated one labeled as QFS$a$.

In order to visualize the main spin-orbital patterns contributing to
the quantum ground state it is convenient to adopt a representation
with arrows for the spin and ellipsoids for the orbital sector at any
given host site. The arrows stand for the on-site spin projection
$\langle S_i^z\rangle$, with the length being proportional to the
amplitude. The length scale for the arrows is the same for all the
configurations. Moreover, in order to describe the orbital character of
the ground state we employed a graphical representation that makes use
of an ellipsoid whose semi-axes $\{a,b,c\}$ length are given by the
average amplitude of the squared angular momentum components
$\{(L^x_i)^2,(L^y_i)^2,(L^z_i)^2\}$, or equivalently by the doublon
occupation Eq. (\ref{eq:dub}). For instance, for a completely flat
circle (degenerate ellipsoid) lying in the plane perpendicular to the
$\gamma$ axis only the corresponding $\gamma$ orbital is occupied.
On the other hand, if the ellipsoid develops in all three directions
$\{a,b,c\}$ it implies that more than one orbital is occupied and the
distribution can be anisotropic in general. If all the orbitals
contribute equally, one finds an isotropic spherical ellipsoid.

Due to the symmetry of the Hamiltonian, the phases shown in the phase
diagram of Fig. \ref{fig:diag}(a) can be characterized by the quantum
numbers for the $z$-th spin projection, $S^z$, and the doublon orbital
occupation $N_{\alpha}$ (\ref{doc}), $\left(S^{z},N_a,N_b,N_c\right)$:
QAF$c1$ $\left(-3.5,2,2,3\right)$, QFS$a2$ $\left(-1.5,3,1,3\right)$,
QAF$a1$ $\left(-5.5,1,3,3\right)$, QAF$a2$ and
QAF$a2$ $\left(-5.5,2,2,3\right)$, QFS$a1$ $\left(-0.5,3,0,4\right)$,
and QFM$a$ $\left(-8.5,2,1,4\right)$. Despite the irregular shape of
the cluster [Fig. \ref{fig:diag}(b)] there is also symmetry between
the $a$ and $b$ directions. For this reason, the phases with
$N_{a}\neq N_{b}$ can be equivalently described either by the set
$\left(S^z,N_a,N_b,N_c\right)$ or $\left(S^z,N_b,N_a,N_c\right)$.

The outcome of the quantum analysis indicates that the spin patterns
are quite robust as the spin configurations of the phases QAF$a$,
QAF$c$, QFS$a$ and QFM$a$ are the analogues of the classical ones. The
effects of quantum fluctuations are more evident in the orbital sector
where mixed orbital patterns occur if compared to the classical case.
In particular, orbital inactive states around the impurity are softened 
by quantum fluctuations and on some bonds we find
an orbital configuration with a superposition of active and inactive
states. The unique AF states where the classical inactive scenario is
recovered corresponds to the QAF$c1$ and QAF$c2$ ones in the regime of
small $\eta_{\rm imp}$. A small hybridization of active and inactive
orbitals along both the AF and FM bonds is also observed around the
impurity for the QFS$a$ phases as one can note by the shape of the
ellipsoid at host sites. Moreover, in the range of large
$\eta_{\rm imp}$ where the FM state is stabilized, the orbital pattern
around the impurity is again like in the classical case.

A significant orbital rearrangement is also obtained within the host.
We generally obtain an orbital pattern that is slightly modified 
from the pure AO configuration assumed in the classical case. The
effect is dramatically different in the regime of strong impurity-host
coupling (i.e., for large $J_{\rm{\rm imp}}$) with AF exchange
(QAF$a2$) with the formation of an orbital liquid around the impurity
and within the host, with doublon occupation represented by an almost 
isotropic shaped ellipsoid. Interestingly, though with a different 
orbital arrangement, the QFS$a1$ and the QFS$a2$ states are the only 
ones where the $C$-AF order of the host is recovered. For all the other 
phases shown in the diagram of Fig. \ref{fig:diag} the coupling between 
the host and the impurity is generally leading to a uniform spin 
polarization with FM or AF coupling between the host and the impurity 
depending on the strength of the host-impurity coupling. Altogether, 
we conclude that the classical spin patterns are only quantitatively 
modified and are robust with respect to quantum fluctuations.

\subsection{Finite spin-orbit coupling}
\label{sec:soc}

In this Section we analyze the quantum effects in the spin and orbital
order around the impurity in the presence of the spin-orbit coupling at
the host $d^4$ sites. For the $t_{2g}^{4}$ configuration the strong
spin-orbit regime has been considered recently by performing an
expansion around the atomic limit where the angular $\vec{L}_i$ and 
spin $\vec{S}_i$ momenta form a spin-orbit singlet for the amplitude of 
the total angular momentum, $\vec{J}_i=\vec{L}_i+\vec{S}_i$ (i.e., $J=0$)
\cite{Kha13}. The instability towards an AF state starting from the 
$J=0$ liquid has been obtained within the spin-wave theory \cite{Kha14} 
for the low energy excitations emerging from the spin-orbital exchange.

In the analysis presented here we proceed from the limit of zero
spin-orbit to investigate how the spin and orbital order are gradually
suppressed when approaching the $J=0$ spin-orbit singlet state. This
issue is addressed by solving the full quantum Hamiltonian (\ref{fullH})
exactly on a cluster of $L=8$ sites including the spin-orbital exchange
for the host and that one derived for the host-impurity coupling
(\ref{fullH}) as well as the spin-orbit term,
\begin{equation}
{\cal H}_{so}=\lambda \sum_{i\in\rm host}\vec{L}_i\cdot\vec{S}_i.
\label{Hso}
\end{equation}
where the sum includes the ions of the $4d$ host and we use the spin
$S=1$ and the angular momentum $L=1$, as in the ionic $4d^4$ 
configurations. Here $\lambda$ is the spin-orbit coupling constant at 
$4d$ host ions, and the components of the orbital momentum
$\vec{L}_i\equiv\{L^x_i,L^y_i,L^z_i\}$ are defined as follows:
\begin{eqnarray}
L^x_i&=&i \sum_{\sigma}(d^{\dagger}_{i,xy\sigma}d_{i,xz\sigma}
                       -d^{\dagger}_{i,xz\sigma}d_{i,xy\sigma}), \nonumber \\
L^y_i&=&i \sum_{\sigma}(d^{\dagger}_{i,xy\sigma}d_{i,yz\sigma}
                       -d^{\dagger}_{i,yz\sigma}d_{i,xy\sigma}), \nonumber \\
L^z_i&=&i \sum_{\sigma}(d^{\dagger}_{i,xz\sigma}d_{i,yz\sigma}
                       -d^{\dagger}_{i,yz\sigma}d_{i,xz\sigma}).
\end{eqnarray}
To determine the ground state and the relevant correlation functions
we use again the Lanczos algorithm for the cluster of $L=8$ sites.
Such an approach allows us to study the competition between the 
spin-orbital exchange and the spin-orbit coupling on equal footing 
without any simplifying approximation. Moreover,
the cluster calculation permits to include the impurity in the host
and deal with the numerous degrees of freedom without making
approximations that would constrain the interplay of the 
impurity-host versus host-host interactions.

Finite spin-orbit coupling significantly modifies the symmetry
properties of the problem. Instead of the SU(2) spin invariance one has
to deal with the rotational invariance related to the total angular
momentum per site $\vec{J}_i=\vec{L}_i+\vec{S}_i$. Though the
$\vec{L}_i\cdot\vec{S}_i$ term in Eq. (\ref{Hso}) commutes with both
total $\vec{J}^2$ and $J^z$, the full Hamiltonian for the host with
impurities Eq. (\ref{fullH}) has a reduced symmetry because the spin
sector is now linearly coupled to the orbital which has only the cubic
symmetry. Thus the remaining symmetry is a cyclic permutation of the
$\{x,y,z\}$ axes.

Moreover, $J^z$ is not a conserved quantity due to the orbital
anisotropy of the spin-orbital exchange in the host and the orbital
character of the impurity-host coupling. There one has a $\mathbb{Z}_2$
symmetry associated with the parity operator (-1)$^{J_{z}}$.
Hence, the ground state can be classified as even or odd with respect
to the value of $J^z$. This symmetry aspect can introduce a constraint
on the character of the ground state and on the impurity-host coupling
since the $J^z$ value for the impurity is only due to the spin
projection while in the host it is due to the combination of the
orbital and spin projection. A direct consequence is that the parity
constraint together with the unbalance between the spin at the host and
the impurity sites leads to a nonvanishing total projection of the spin
and angular momentum with respect to a symmetry axis, e.g. the $z^{th}$ 
axis. It is worth to note that a fixed parity for the impurity spin 
means that it prefers to point in one direction rather than the other 
one which is not the case for the host's spin and angular momentum. Thus 
the presence on the impurity for a fixed ${\cal P}$ will give a nonzero
polarization along the quantization axis for every site of the system.
Such a property holds for any single impurity with a half-integer spin.

Another important consequence of the spin-orbit coupling is that it
introduces local quantum fluctuations in the orbital sector even at
the sites close to the impurity where the orbital pattern is disturbed.
The spin-orbit term makes the on-site problem around the impurity
effectively analogous to the Ising model in a transverse field for the
orbital sector, with nontrivial spin-orbital entanglement \cite{Ole12} 
extending over the impurity neighborhood.

\begin{figure*}[t!]
\begin{centering}
\includegraphics[width=0.92\textwidth]{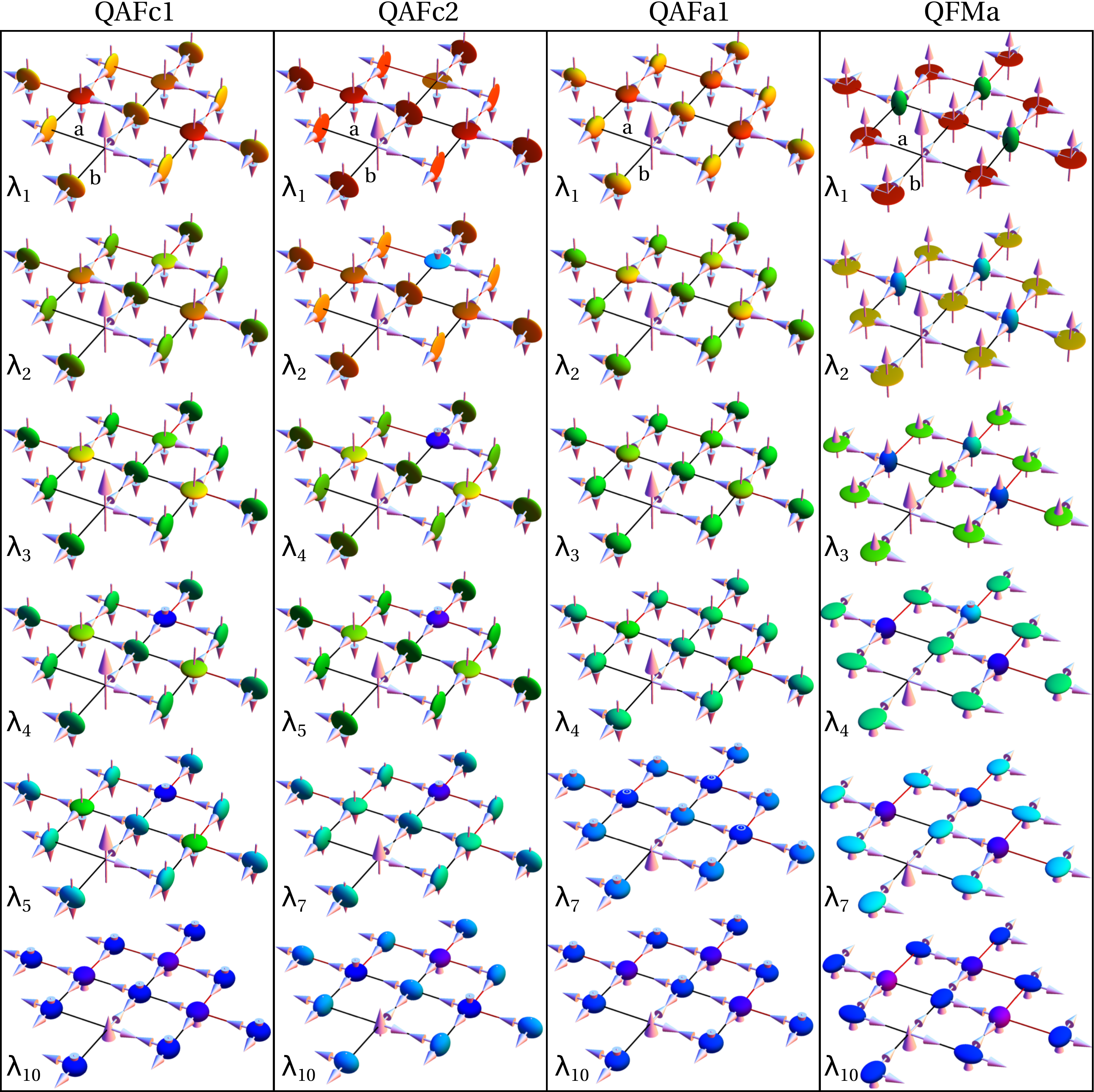}
\par\end{centering}
\protect\caption{
Evolution of the ground state configurations for the AF phases for
selected increasing values of spin-orbit coupling $\lambda_m$,
see Eq. (\ref{lambda}).
Arrows and ellipsoids indicate the spin-orbital state at a given site
$i$. Color map indicates the strength of the average spin-orbit,
$\langle\vec{L}_i \cdot \vec{S}_i\rangle$, i.e., red, yellow, green, blue,
violet correspond to the growing amplitude of the above correlation
function.
Small arrows at $\lambda_5$ and $\lambda_{10}$ indicate quenched
magnetization at the impurity by large spin-orbit coupling at the
neighboring host sites.}
\label{fig:4conf}
\end{figure*}

\begin{figure}[t!]
\begin{centering}
\includegraphics[width=0.95\columnwidth]{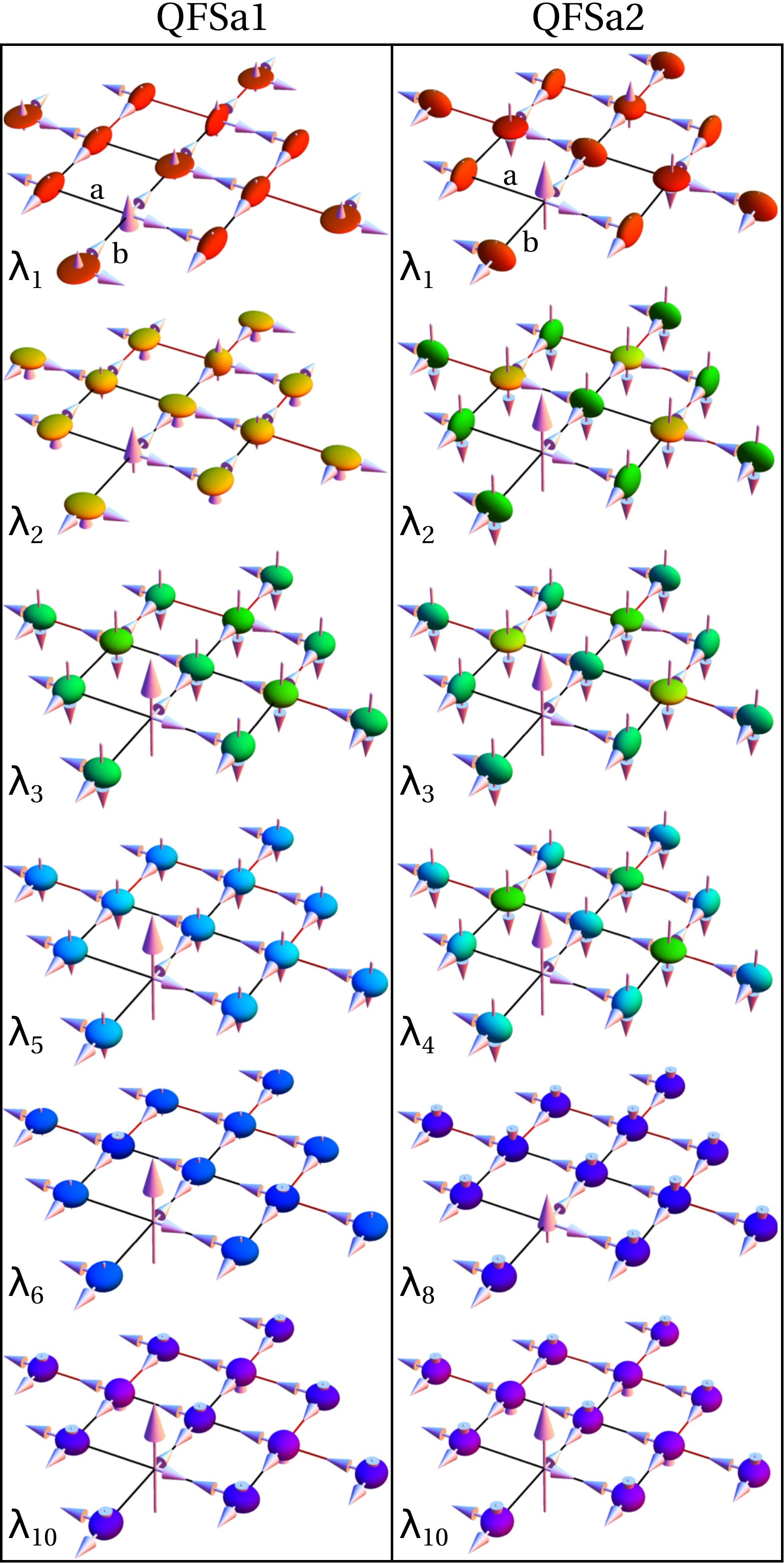}
\par\end{centering}
\protect\caption{
Evolution of the ground state configurations for the QFS$a1$ and QFS$a2$ 
phases for selected increasing values of spin-orbit coupling $\lambda_m$, 
see Eq. (\ref{lambda}).
Arrows and ellipsoids indicate the spin-orbital state at a given site
$i$. Color map indicates the strength of the average spin-orbit,
$\langle{\vec L}_i\cdot{\vec S}_i\rangle$, i.e., red, yellow, green,
blue, violet correspond to the growing amplitude of the above local
correlation function.}
\label{fig:2conf}
\end{figure}

In Figs. \ref{fig:4conf} and \ref{fig:2conf} we report the schematic 
evolution of the ground state configurations for the cluster of 
$L=8$ sites, with one-impurity and periodic boundary conditions as a
function of increasing spin-orbit coupling. These patterns have been
determined by taking into account the sign and the amplitude of the
relevant spatial dependent spin and orbital correlation functions.
The arrows associated to the spin degree of freedom can lie in $xy$
plane or out-of-plane (along $z$, chosen to be parallel to the
$c$ axis) to indicate the anisotropic spin pattern. The out-of-plane
arrow length is given by the on-site expectation value of
$\langle S^z_i\rangle$ while the in-plane arrow length is obtained by
computing the square root of the second moment, i.e.,
$\sqrt{\langle(S^x_i)^2}\rangle$ and $\sqrt{\langle(S^y_i)^2}\rangle$
of the $x$ and $y$ spin components corresponding to the arrows along
$a$ and $b$, respectively.

Moreover, the in-plane arrow orientation for a given direction is
determined by the sign of the corresponding spin-spin correlation
function assuming as a reference the orientation of the impurity spin.
The ellipsoid is constructed in the same way as for the zero spin-orbit
case above, with the addition of a color map that indicates the
strength of the average $\vec{L}_i\cdot\vec{S}_i$ (i.e., red, yellow,
green, blue, violet correspond with a growing amplitude of the local
spin-orbit correlation function). The scale for the spin-orbit
amplitude is set to be in the interval $0<\lambda<J_{\rm host}$.
The selected values for the ground state evolution are given by the
relation (with $m=1,2,\dots,10$),
\begin{equation}
\lambda_m=\left[\,0.04+0.96\,\frac{(m-1)}{9}\,\right]J_{\rm host}.
\label{lambda}
\end{equation}
The scale is set such that $\lambda_1=0.04J_{\rm host}$ and
$\lambda_{10}=J_{\rm host}$. This range of values allows us to explore
the relevant physical regimes when moving from $3d$ to $4d$ and $5d$
materials with corresponding $\lambda$ being much smaller that
$J_{\rm host}$, $\lambda\sim J_{\rm host}/2$ and $\lambda>J_{\rm host}$,
respectively. For the performed analysis the selected values of 
$\lambda$ (\ref{lambda}) are also representative of the most 
interesting regimes of the ground state as induced by the spin-orbit 
coupling.

Let us start with the quantum AF phases QAF$c1$, QAF$c2$, QFS$a1$, 
QFS$a2$, QAF$a1$, and QAF$a2$. As one can observe the switching on of 
the spin-orbit coupling (i.e., $\lambda_1$ in Fig. \ref{fig:4conf}) 
leads to anisotropic spin patterns with unequal moments for the 
in-plane and out-of-plane components.
From weak to strong spin-orbit coupling, the character of the spin
correlations keeps being AF between the impurity and the neighboring
host sites in all the spin directions. The main change for the spin
sector occurs for the planar components. For weak spin-orbit coupling
the in-plane spin pattern is generally AF for the whole system in all
the spatial directions (i.e., $G$-AF order). Further increase of
the spin-orbit does not modify qualitatively the character of the spin
pattern for the out-of-plane components as long as we do not go to
maximal values of $\lambda\sim J_{\rm host}$ where local
$\langle S_i^z\rangle$ moments are strongly suppressed. In this
limit the dominant tendency of the system is towards formation of
the spin-orbital singlets and the spin patterns shown in Fig.
\ref{fig:4conf} are the effect of the virtual singlet-triplet
excitations \cite{Kha13}.

Concerning the orbital sector, only for weak spin-orbit coupling around
the impurity one can still observe a reminiscence of inactive orbitals
as related to the orbital vacancy role at the impurity site in the AF
phase. Such an orbital configuration is quickly modified by increasing
the spin-orbit interaction and it evolves into a uniform pattern with
almost degenerate orbital occupations in all the directions, and with
preferential superpositions of $c$ and $(a,b)$ states associated with
dominating $L^x$ and $L^y$ orbital angular components (flattened
ellipsoids along the $c$ direction). An exception is the QAF$c2$ phase
with the orbital inactive polaron that is stable up to large spin-orbit
coupling of the order of $J_{\rm host}$.

When considering the quantum FM configurations QFM$a1$ in Fig. 
\ref{fig:4conf}, we observe similar trends in the evolution of the spin 
correlation functions as obtained for the AF states. Indeed, the QFM$a$ 
exhibits a tendency to form FM chains with AF coupling for the in-plane 
components at weak spin-orbit that evolve into more dominant AF 
correlations in all the spatial directions within the host. 
Interestingly, the spin exchange between the impurity and the 
neighboring host sites shows a changeover from AF to FM for the range 
of intermediate-to-strong spin-orbit amplitudes.

A peculiar response to the spin-orbit coupling is obtained for the
QFS$a1$ phase, see Fig. \ref{fig:2conf}, which showed a frustrated spin 
pattern around the impurity already in the classical regime, with FM and 
AF bonds. It is remarkable that due to the close proximity with uniform 
FM and the AF states, the spin-orbit interaction can lead to a dramatic 
rearrangement of the spin and orbital correlations for such a 
configuration. At weak spin-orbit coupling (i.e., 
$\lambda\simeq\lambda_1$) the spin-pattern is $C$-AF and the increased 
coupling ($\lambda\simeq\lambda_2)$) keeps the $C$-AF order only for the 
in-plane components with the exception of the impurity site. It also 
modulates the spin moment distribution around the impurity along the $z$ 
direction. Further increase of $\lambda$ leads to complete spin 
polarization along the $z$ direction in the host, with antiparallel 
orientation with respect to the impurity spin. This pattern is guided by 
the proximity to the FM phase. The in-plane components develop a mixed 
FM-AF pattern with a strong $xy$ anisotropy most probably related to the 
different bond exchange between the impurity and the host.

When approaching the regime of a spin-orbit coupling that is comparable
to $J_{\rm{host}}$, the out-of-plane spin components dominate and the
only out-of-plane spin polarization is observed at the impurity site.
Such a behavior is unique and occurs only in the QFS$a$ phases.
The cooperation between the strong spin-orbit coupling and the
frustrated host-impurity spin-orbital exchange leads to an effective
decoupling in the spin sector at the impurity with a resulting maximal
polarization. On the other hand, as for the AF states, the most
favorable configuration for strong spin-orbit has AF in-plane spin
correlations. The orbital pattern for the QFS$a$ states evolves 
similarly to the AF cases with a suppression of the active-inactive 
interplay around the impurity and the setting of a uniform-like orbital
configuration with unquenched angular momentum on site and predominant
in-plane components. The response of the FM state is different in this
respect as the orbital active states around the impurity are hardly
affected by the spin-orbit while the host sites far from the impurity
the local spin-orbit coupling is more pronounced.

Finally, to understand the peculiar evolution of the spin configuration
it is useful to consider the lowest order terms in the spin-orbital
exchange that couple directly the orbital angular momentum with the spin.
Taking into account the expression of the spin-orbital exchange in the
host (\ref{Hso}) and the expression of $\vec{L}_i$ one can show that the 
low energy terms on a bond that get more relevant in the Hamiltonian 
when the spin-orbit coupling makes a non-vanishing local angular 
momentum. As a result, the corresponding expressions are:
\begin{eqnarray}
H_{\rm host}^{a(b)}(i,j)&\approx &
J_{{\rm host}}\left\{ a_1\vec{S}_{i}\!\cdot\!\vec{S}_{j}+ b_1S^{z}_{i} S^{z}_{j}L_{i}^{y(x)}L_{j}^{y(x)}\right\}  \nonumber \\
&+&\lambda\left\{ \vec{L}_{i}\!\cdot\!\vec{S}_{i}
+\vec{L}_{j}\!\cdot\!\vec{S}_{j}\right\},
\label{ls}
\end{eqnarray}
with positive coefficients $a_1$ and $b_1$ that depend on $r_1$ and 
$r_2$ (\ref{rr}). A definite sign for the spin exchange in the limit of
vanishing spin-orbit coupling is given by the terms which go beyond Eq.
(\ref{ls}). Then, if the ground state has isotropic FM correlations
(e.g. QFM$a$) at $\lambda=0$, the term $S^z_iS^z_jL_i^{y(x)}L_j^{y(x)}$
would tend to favor AF-like configurations for the in-plane
orbital angular components when the spin-orbit interaction is switched
on. This opposite tendency between the $z$ and $\{x,y\}$ components is
counteracted by the local spin-orbit coupling that prevents to have
coexisting FM and AF spin-orbital correlations. Such patterns would not
allow to optimize the $\vec{L}_i\cdot\vec{S}_i$ amplitudes. One way out
is to reduce the $z^{th}$ spin projection and to get planar AF 
correlations in the spin and in the host. A similar reasoning applies 
to the AF states where the negative sign of the $S^{z}_iS^{z}_j$ 
correlations favors FO alignment of the angular momentum components. 
As for the previous case, the opposite trend of in- and out-of-plane 
spin-orbital components is suppressed by the spin-orbit coupling and 
the in-plane FO correlations for the $\{L^x,L^y\}$ components leads to 
FM patterns for the in-plane spin part as well.

Summarizing, by close inspection of Figs. \ref{fig:4conf} and
\ref{fig:2conf} one finds an interesting evolution of the spin patterns
in the quantum phases:
\\ \noindent
(i) For the QAF states (Fig. \ref{fig:4conf}),
a spin canting develops at the host sites (i.e., the
relative angle is between 0 and $\pi$) while the spins on impurity-host
bonds are always AF. The canting in the host evolves, sometime in an
inhomogeneous way, to become reduced in the strong spin-orbit coupling
regime where ferro-like correlations tend to dominate. In this respect,
when the impurity is coupled antiferromagnetically to the host it does
not follow the tendency to form spin canting.
\\ \noindent
(ii) In the QFM states (Fig. \ref{fig:2conf}), at weak spin-orbit one 
observes spin-canting in the host and for the host-impurity coupling 
that persists only in the host whereas the spin-orbit interaction is
increasing.

\subsection{Spin-orbit coupling versus Hund's exchange}
\label{sec:JH}

\begin{figure}[t!]
\begin{centering}
\includegraphics[width=.99\columnwidth]{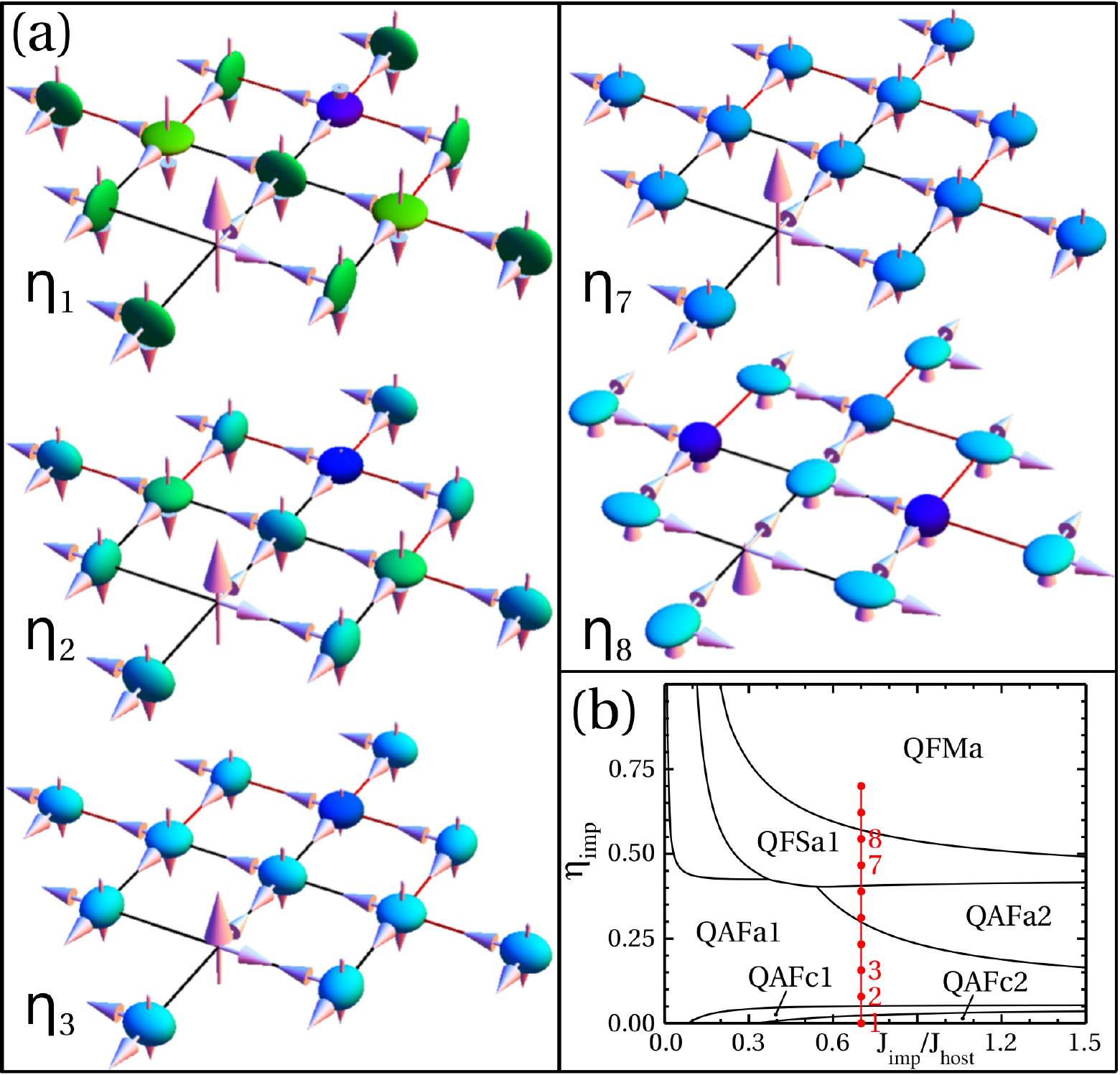}
\par\end{centering}
\protect\caption{(a) Evolution of the ground state configurations as
for increasing $\eta_{\rm imp}$ and for a fixed value of spin-orbit 
coupling $\lambda=0.5J_{\rm host}$ along a cut in the phase diagram 
shown in panel (b), i.e., for 
$J_{\rm imp}=0.7J_{\rm host}$ and $0\leq\eta_{\rm imp}\leq 0.7$.
Arrows and ellipsoids indicate the spin-orbital state at a given site 
$i$. Color map indicates the strength of the average spin-orbit,
$\langle\vec{L}_i\cdot\vec{S}_i\rangle$, i.e., red, yellow, green, 
blue, violet correspond to the growing amplitude of the above 
correlation function.}
\label{fig:JH}
\end{figure}

To probe the phase diagram of the system in presence of the spin-orbit
coupling ($\lambda>0$) we solved the same cluster of $L=8$ sites as 
before along three different cuts in the phase diagram of 
Fig. \ref{fig:diag}(a) for three values of $\lambda$, i.e., 
small $\lambda=0.1J_{\rm host}$,
intermediate $\lambda=0.5J_{\rm host}$, and 
large $\lambda=J_{\rm host}$.
Each cut contained ten points, the cuts were parameterized as follows:
(i)  $J_{\rm imp}=0.7J_{\rm host}$ and $0\leq\eta_{\rm imp}\leq0.7$,
(ii) $J_{\rm imp}=1.3J_{\rm host}$ and $0\leq\eta_{\rm imp}\leq0.7$,
and
(iii) $\eta_{\rm imp}=\eta_{\rm imp}^{c}\simeq0.43$ and
$0\leq J_{\rm imp}\leq1.5J_{\rm host}$.
In Fig. \ref{fig:JH}(a) we show the representative spin-orbital
configurations obtained for $\lambda=0.5J_{\rm host}$ along the first
cut shown in Fig. \ref{fig:JH}(b). Values of $\eta_{\rm imp}$ are
chosen as
\begin{equation}
\eta_{\rm imp}=\eta_{m}\equiv 0.7\,\frac{(m-1)}{9},
\label{cut}
\end{equation}
with $m=1,\dots,10$ but not all the points are shown in Fig.
\ref{fig:JH}(a) --- only the ones for which the spin-orbital
configuration changes substantially.

The cut starts in the QAF$c2$ phase, according to the phase diagram 
of Fig. \ref{fig:JH}(b), and indeed we find a similar configuration
to the one shown in Fig. \ref{fig:4conf} for QAF$c2$ phase at
$\lambda=\lambda_5$. Moving up in the phase diagram from $\eta_1$ to 
$\eta_2$ we see that the configuration evolves smoothly to the one 
which we have found in the QAF$a1$ phase at $\lambda=\lambda_5$ 
(not shown in Fig. \ref{fig:4conf}). 
The evolution of spins is such that the out-of-plane
moments are suppressed while in-plane ones are slightly enhanced. 
The orbitals become more spherical and the local spin-orbit average,
$\langle \vec{L}_i\cdot\vec{S}_i\rangle$,  becomes larger and more 
uniform, however for the apical site $i=7$ in the cluster 
[Fig. \ref{fig:diag}(b)] the trend is opposite --- initially large 
value of spin-orbit coupling drops towards the uniform value. The 
points between $\eta_3$ and $\eta_7$ we skip as the evolution is 
smooth and the trend is clear, however the impurity out-of-plane 
moment begins to grow above $\eta_5$, indicating proximity to the 
QFS$a1$ phase. For this phase at intermediate and high $\lambda$ 
the impurity moment is much larger than all the others 
(see Fig. \ref{fig:2conf}).

For $\eta_{\rm imp}=\eta_7$ the orbital pattern clearly shows that we 
are in the QFS$a1$ phase at $\lambda=\lambda_5$ which agrees with the 
position of the $\eta_7$ point in the phase diagram, see Fig.
\ref{fig:JH}(b). On the other hand, moving to the next $\eta_{\rm imp}$ 
point upward along the cut Eq. (\ref{cut}) we already observe a 
configuration which is very typical for the QFM$a$ phase at 
intermediate $\lambda$ (here $\lambda=\lambda_7$ shown in Fig. 
\ref{fig:4conf} but also $\lambda_6$, not shown). 
This indicates that the QFS$a1$ phase can be still distinguished at 
$\lambda=0.5J_{\rm host}$ and its position in the phase diagram is
similar as in the $\lambda=0$ case, i.e., as an intermediate phase
between the QAF$a1(2)$ and QFM$a$ one.

Finally, we have found that also the two other cuts which were
not shown here, i.e., for $J_{\rm imp}=1.3J_{\rm host}$ and 
increasing $\eta_{\rm imp}$ and for 
$\eta_{\rm imp}=\eta_{\rm imp}^{c}\simeq0.43$ and increasing 
$J_{\rm imp}$ confirm that the overall character of the phase
diagram of Fig. \ref{fig:diag}(a) is preserved at this value of
spin-orbit coupling, however firstly, the transitions between the
phases are smooth and secondly, the subtle differences between the 
two QFS$a$, QAF$a$ and QAF$c$ phases are no longer present. 
This also refers to the smaller value of $\lambda$, i.e.,
$\lambda=0.1J_{\rm host}$, but already for $\lambda=J_{\rm host}$ the
out-of-plane moments are so strongly suppressed (except for the impurity
moment in the QFS$a1$ phase) and the orbital polarization is so weak
(i.e., almost spherical ellipsoids) that typically the only distinction
between the phases can be made by looking at the in-plane spin
correlations and the average spin-orbit,
$\langle\vec{L}_i\cdot\vec{S}_i\rangle$. In this limit we conclude that
the phase diagram is (partially) melted by large spin-orbit coupling
but for lower values of $\lambda$ it is still valid.

\section{Summary and conclusions}
\label{sec:sum}

We have derived the spin-orbital superexchange model for $3d^3$ 
impurities replacing $4d^4$ (or $3d^2$) ions in the $4d$ ($3d$) host 
in the regime of Mott insulating phase. Although the impurity has no 
orbital degree of freedom, we have shown that it contributes to the 
spin-orbital physics and influences strongly the orbital order. In 
fact, it tends to project out the inactive orbitals at the
impurity-host bonds to maximize the energy gain from virtual charge
fluctuations. In this case the interaction along the superexchange 
bond can be either antiferromagnetic or ferromagnetic, depending on
the ratio of Hund's exchange coupling at impurity ($J_1^H$) and host
($J_2^H$) ions and on the mismatch $\Delta$ between the $3d$ and $4d$
atomic energies, modified by the difference in Hubbard $U$'s and Hund's 
exchange $J^H$'s at both atoms. This ratio, denoted $\eta_{\rm imp}$ 
(\ref{eq:etai}), replaces here the conventional parameter 
$\eta=J_H/U$ often found in the spin-orbital superexchange models of
undoped compounds (e.g., in the Kugel-Khomskii model for KCuF$_3$
\cite{Fei97}) where it quantifies the proximity to ferromagnetism. On 
the other hand, if the overall coupling between the host and impurity
is weak in the sense of the total superexchange, $J_{\rm imp}$, with
respect to the host value, $J_{\rm host}$, the orbitals being next
to the impurity may be forced to stay inactive which modifies the
magnetic properties --- in such cases the impurity-host bond is
always antiferromagnetic.

As we have seen in the case of a single impurity, the above two
mechanisms can have a nontrivial effect on the host, especially if
the host itself is characterized by frustrated interactions, as it
happens in the parameter regime where the $C$-AF phase is stable.
For this reason we have focused mostly on the latter phase of the
host and we have presented the phase diagrams of a single impurity
configuration in the case when the impurity is doped on the sublattice
where the orbitals form a checkerboard pattern with alternating $c$ and
$a$ orbitals occupied by doublons. The diagram for the $c$-sublattice
doping shows that in some sense the impurity is never weak, because
even for a very small value of $J_{\rm imp}/J_{\rm host}$ it can
release the host's frustration around the impurity site acting as an
orbital vacancy. On the other hand, for the $a$-sublattice doping when
the impurity-host coupling is weak, i.e., either
$J_{\rm imp}/J_{\rm host}$ is weak or $\eta_{\rm imp}$ is close to
$\eta_{\rm imp}^c$, we have identified an interesting quantum mechanism
releasing frustration of the impurity spin (that cannot be avoided in
the purely classical approach). It turned out that in such situations
the orbital flips in the host make the impurity spin polarize in such
a way that the $C$-AF order of the host is completely restored.

The cases of the periodic doping studied in this paper show that the
host's order can be completely altered already for \textit{rather low}
doping of $x=1/8$, even if the $J_{\rm imp}/J_{\rm host}$ is small.
In this case we can stabilize a ferrimagnetic type of phase with a
four-site unit cell having magnetization $\langle S_i^z\rangle=3/2$,
reduced further by quantum fluctuations.
We have established that the only parameter range where the
host's order remains unchanged is when $\eta_{\rm imp}$ is close to
$\eta_{\rm imp}^c$ and $J_{\rm imp}/J_{\rm host}\gtrsim 1$. The latter
value is very surprising as it means that the impurity-host coupling
must be large enough to keep the host's order unchanged --- this is
another manifestation of the orbital vacancy mechanism that we
have already observed for a single impurity. Also in this case
the impurity spins are fixed with the help of orbital flips in the 
host that lift the degeneracy which arises in the classical approach.
We would like to point out that the quantum
mechanism that lifts the ground state degeneracy mentioned above and
the role of quantum fluctuations are of particular interest for the
periodically doped checkerboard systems with $x=1/2$ doping which
is a challenging problem for future research.

From the point of view of generic, i.e., non-periodic doping, the most
representative cases are those of a doping which is incommensurate with 
the two-sublattice spin-orbital pattern. To uncover the generic rules 
in such cases, we have studied periodic $x=1/5$ and $x=1/9$ doping. 
One finds that when the period of the impurity positions does not match 
the period of 2 for both the spin and orbital order of the host, 
interesting novel types of order emerge. In such cases the elementary 
cell must be doubled in both lattice directions which clearly gives 
a chance of realizing more phases than in the case of commensurate 
doping. Our results show that indeed, the number of phases increases 
from 4 to 7 and the host's order is altered in each of them. Quite 
surprisingly, the overall character of the phase diagram remained 
unchanged with respect to the one for $x=1/8$ doping and, if we ignore 
the differences in configuration, it seems that only some of the
phases got divided into two versions differing either by the spin
bond's polarizations around impurities (phases around 
$\eta_{\rm imp}^c$), or by the character of the orbitals around the
impurities (phases with inactive orbitals in the limit of small
enough product $\eta_{\rm imp}J_{\rm imp}$, versus phases with active
orbitals in the opposite limit).
Orbital polarization in this latter region resembles orbital polarons
in doped manganites \cite{Dag04,Gec05} --- also here such states are
stabilized by the double exchange \cite{deG60}.

A closer inspection of
underlying phases reveals however a very interesting degeneracy of the
impurity spins at $x=1/5$ that arises again from the classical approach
but this time it cannot be released by short-range orbital flips. This
happens because the host's order is already so strongly altered that
it is no longer anisotropic (as it was the case of the $C$-AF phase)
and there is no way to restore the orbital anisotropy around the
impurities that could lead to spin-bonds imbalance and polarize the
spin. In the case of lower $x=1/9$ doping such an effect is absent and 
the impurity spins are always polarized, as it happens for $x=1/8$. 
It shows that this is rather a peculiarity of the $x=1/5$ periodic 
doping. 

Indeed, one can easily notice that for $x=1/5$ every atom of the host 
is a nearest neighbor of some impurity. In contrast, for $x=1/8$ we 
can find three host's atoms per unit cell which do not neighbor any 
impurity and for $x=1/9$ there are sixteen of them. For this reason 
the impurity effects are amplified for $x=1/5$ which is not unexpected 
although one may find somewhat surprising that the ground state diagrams 
for the lowest and the highest doping considered here are very similar. 
This suggests that the cooperative effects of multiple impurities are 
indeed not very strong in the low-doping regime, so the diagram obtained 
for $x=1/9$ can be regarded as generic for the dilute doping regime with 
uniform spatial profile.

For the representative case of $x=1/8$ doping, we have presented the
consequences of quantum effects beyond the classical approach. Spin
fluctuations are rather weak for the considered case of large $S=1$ 
and $S=3/2$ spins, and we have shown that orbital fluctuations on
superexchange bonds are more important. They are strongest in the
regime of antiferromagnetic impurity-host coupling (which suggests
importance of entangled states \cite{Ole12}) and enhance the tendency
towards frustrated impurity spin configurations but do not destroy 
other generic trends observed when the parameters $\eta_{\rm imp}$ 
and $J_{\rm imp}/J_{\rm host}$ increase.

Increasing spin-orbit coupling leads to qualitative changes in the
spin-orbital order. When Hund's exchange is small at the impurity
sites, the antiferromagnetic bonds around it have reduced values of
spin-orbit coupling term, but the magnetic moments reorient and survive 
in the $(a,b)$ planes, with some similarity to the phenomena occurring 
in the perovskite vanadates \cite{Hor03}. This quenches the magnetic 
moments at $3d$ impurities and leads to almost uniform orbital 
occupancies at the host sites. In contrast, frustration of impurity 
spins is removed and the impurity magnetization along the $c$ axis 
survives for large spin-orbit coupling.

We would like to emphasize that the \textit{orbital dilution} considered
here influences directly the orbital degrees of freedom in the host
around the impurities. The synthesis of hybrid compounds having both
$3d$ and $4d$ transition metal ions will likely open a novel route for
unconventional effects in complex materials. There are several reasons
for expecting new scenarios in mixed $3d-4d$ spin-orbital-lattice
materials, and we pointed out only some of them.
On the experimental side, the changes of local order could be captured
using inelastic neutron scattering or resonant inelastic x-ray
scattering (RIXS). In fact, using RIXS can also bring an additional
advantage: RIXS, besides being a perfect probe of both spin and orbital
excitations, can also (indirectly) detect the nature of orbital ground
state (supposedly also including the nature of impurities in the
crystal) \cite{Woh12}. Unfortunately, there are no such experiments
yet but we believe that they will be available soon.

Short range order around impurities could be investigated by the
excitation spectra at the resonant edges of the substituting atoms.
Taking them both at finite energy and momentum can dive insights into
the nature of the short range order around the impurity and then
unveil information of the order within the host as well. Even if 
there are no elastic superlattice extra peaks one can expect that the
spin-orbital correlations will emerge in the integrated RIXS spectra
providing information of the impurity-host coupling and of the short 
range order around the impurity. Even more interesting is the case 
where the substituting atom forms a periodic array with small deviation 
from the perfect superlattice when one expects the emergence of extra 
elastic peaks which will clearly indicate the spin-orbital 
reconstruction. In our case an active orbital diluted site cannot
participate coherently in the host spin-orbital order but rather may
to restructure the host ordering \cite{Hos13}. At dilute impurity
concentration we may expect broad peaks emerging at finite momenta in
the Brillouin zone, indicating the formation of coherent islands with
short range order around impurities. 

We also note that local susceptibility can be suitably measured by 
making use of resonant spectroscopies (e.g. nuclear magnetic resonance 
(NMR), electron spin resonance (ESR), nuclear quadrupole resonance 
(NQR), muon spin resonance ($\mu$SR), \textit{etcetera}) 
that exploit the different magnetic or electric character of the 
atomic nuclei for the impurity and the host in the hybrid system. 
Finally, the random implantation of the muons in the sample can 
provide information of the relaxation time in different domains 
with unequal dopant concentration which may be nonuniform.
For the given problem the differences 
in the resonant response can give relevant information about the 
distribution of the local fields, the occurrence of local order 
and provide access to the dynamical response within doped domains.
The use of local spectroscopic resonance methods has been widely
demonstrated to be successful when probing the nature and the evolution 
of the ground state in the presence of spin vacancies both for ordered 
and disordered magnetic configurations \cite{Lim02,Bob09,Sen11,Bon12}.

In summary, this study highlights the role of spin defects which lead 
to orbital dilution in spin-orbital systems. Using an example of $3d^3$ 
impurities in a $4d^4$ (or $3d^2$) host we have shown that impurities 
change radically the spin-orbital order around them, independently of 
the parameter regime. As a general feature we have found that doped 
$3d^3$ ions within the host with spin-orbital order have frustrated 
spins and polarize the orbitals of the host when the impurity-host 
exchange as well as Hund's exchange at the impurity are both 
sufficiently large. This remarkable trend is independent of doping and 
is expected to lead to global changes of spin-orbital order in doped 
materials. While the latter effect is robust, we argue that the 
long-range spin fluctuations resulting from the translational 
invariance of the system will likely prevent the ground state from 
being macroscopically degenerate, so if the impurity spins in one unit 
cell happens to choose its polarization then the others will follow.
On the contrary, in the regime of weak Hund's exchange $3d^3$ ions act
not only as spin defects which order antiferromagnetically with respect
to their neighbors, but also induce doublons in inactive orbitals.

Finally,
we remark that this behavior with switching between inactive and
active orbitals by an orbitally neutral impurity may lead to multiple
interesting phenomena at macroscopic doping when global modifications
of the spin-orbital order are expected to occur. Most of the results
were obtained in the classical approximation but we have shown that
modifications due to spin-orbit coupling do not change the main
conclusion. We note that this generic treatment and the general
questions addressed here, such as the release of frustration for
competing spin structures due to periodic impurities, are relevant to
double perovskites \cite{Pau13}. While the local orbital polarization
should be similar, it is challenging to investigate disordered
impurities, both theoretically and in experiment, to find out whether
their influence on the global spin-orbital order in the host is 
equally strong.

\acknowledgments

We thank Maria Daghofer and Krzysztof Wohlfeld for insightful
discussions. W. B. and A. M. O. kindly acknowledge support by the Polish
National Science Center (NCN) under Project No. 2012/04/A/ST3/00331.
W. B. was also supported by the Foundation for Polish Science (FNP)
within the START program.
M. C. acknowledges funding from the EU --- FP7/2007-2013 under
Grant Agreement No. 264098 --- MAMA.

\appendix

\section{Derivation of \MakeLowercase{$3d-4d$} superexchange}

Here we present the details of the derivation of the low energy
spin-orbital Hamiltonian for the $3d^3-4d^4$ bonds around the impurity
at site $i$. ${\cal H}_{3d-4d}(i)$, which follows from the perfurbation 
theory, as given in Eq. (\ref{eq:pert_exp}). Here we consider a single 
$3d^3-4d^4$ bond $\langle ij\rangle$. Two contributions to the 
effective Hamiltonian follow from charge excitations:
(i) ${\cal H}_{J,43}^{(\gamma)}(i,j)$ due to
$d^3_id^4_j\leftrightharpoons d^4_id^3_j$, and
(ii) ${\cal H}_{J,25}^{(\gamma)}(i,j)$ due to
$d^3_id^4_j\leftrightharpoons d^2_id^5_j$. Therefore the low energy
Hamiltonian is,
\begin{equation}
{\cal H}_J^{(\gamma)}(i,j)=
{\cal H}_{J,43}^{(\gamma)}(i,j)+{\cal H}_{J,25}^{(\gamma)}(i,j).
\label{eq:HJ}
\end{equation}

Consider first the processes which conserve the number of doubly
occupied orbitals, $d^3_id^4_j\leftrightharpoons d^4_id^3_j$. Then
by means of spin and orbital projectors, it is possible to express
${\cal H}_{J,43}^{(\gamma)}(i,j)$ for $i=1$ and $j=2$ as 
\begin{eqnarray}
& &{\cal H}_{J,43}^{(\gamma)}(1,2)=
\nonumber \\
&&- \left(\vec{S}_{1}\!\cdot\!\vec{S}_{2}\right)\frac{t^{2}}{18}
 \left\{ \frac{4}{\Delta}-\frac{7}{\Delta+3J_{2}^{H}}
 -\frac{3}{\Delta+5J_{2}^{H}}\right\} \nonumber \\
&&+ D_2^{(\gamma)}\left(\vec{S}_1\!\cdot\!\vec{S}_2\right)
\frac{t^{2}}{18}\left\{ \frac{4}{\Delta}-\frac{1}{\Delta+3J_{2}^{H}}
+\frac{3}{\Delta+5J_{2}^{H}}\right\} \nonumber \\
&&+ \left(D_2^{(\gamma)}\!-1\right)\frac{t^{2}}{12}
\left\{ \frac{8}{\Delta}+\frac{1}{\Delta+3J_{2}^{H}}
-\frac{3}{\Delta+5J_{2}^{H}}\right\},
\label{eq:s-ex}
\end{eqnarray}
with the excitation energy $\Delta$ defined in Eq. (\ref{Delta}).
The resulting effective $3d-4d$ exchange in Eq. (\ref{eq:s-ex})
consists of three terms:
(i) The first one does not depend on the orbital configuration of the
$4d$ atom and it can be FM or AF depending on the values $\Delta$ and 
the Hund's exchange on the $3d$ ion. In particular, if $\Delta$ is the 
largest or the smallest energy scale, the coupling will be either AF 
or FM, respectively.
(ii) The second term has an explicit dependence on the occupation of
the doublon on the $4d$ atom via the projecting operator 
$D_2^{(\gamma)}$. This implies that a magnetic exchange is possible
only if the doublon occupies the inactive orbital for a bond along a
given direction $\gamma$. Unlike in the first term, the sign of this 
interaction is always positive favoring an AF configuration at any
strength of $\Delta$ and $J_{1}^{H}$.
(iii) Finally, the last term describes the effective processes which do
not depend on the spin states on the $3d$ and $4d$ atoms. 
This contribution is of pure orbital nature, as it originates from the 
hopping between $3d$ and $4d$ atoms without affecting their spin 
configuration, and for this reason favors the occupation of active 
$t_{2g}$ orbitals along the bond by the doublon.

Within the same scheme, we have derived the effective spin-orbital
exchange that originates from the charge transfer processes of the
type $3d_1^3 4d_2^4\leftrightharpoons 3d_i^24d_j^5$, 
${\cal H}_{J,25}^{(\gamma)}(1,2)$. The effective low-energy 
contribution to the Hamiltonian for $i=1$ and $j=2$ reads
\begin{eqnarray}
\label{eq:s-ex2}
&&{\cal H}_{J,25}^{(\gamma)}(1,2)=
\frac{t^{2}}{U_{1}+U_{2}-\left(\Delta+3J_{2}^{H}-2J_{1}^{H}\right)}
\nonumber\\
&\!\times\!&\left\{\frac{1}{3}D_2^{(\gamma)}\!
\left(\vec{S}_{1}\!\cdot\!\vec{S}_{2}\right)
\!+\frac{1}{3}\!\left(\vec{S}_{1}\!\cdot\!\vec{S}_{2}\right)
\!-\frac{1}{2}\!\left(D_2^{(\gamma)}\!+1\right)\!\right\}.
\end{eqnarray}
By inspection of the spin structure involved in the elemental 
processes that generate ${\cal H}_{J,25}^{(\gamma)}(1,2)$, one can note 
that it is always AF independently of the orbital configuration on the 
$4d$ atom exhibiting with a larger spin-exchange and an orbital energy 
gain if the doublon is occupying the inactive orbital along a given 
bond. We have verified that the amplitude of the exchange terms in 
${\cal H}_{J,25}^{(\gamma)}(1,2)$ is much smaller than the ones which 
enter in ${\cal H}_{J,43}^{(\gamma)}(1,2)$ which justifies that one may 
simplify Eq. (\ref{eq:HJ}) for $i=1$ and $j=2$ to
\begin{equation}
{\cal H}_J^{(\gamma)}(1,2)\simeq {\cal H}_{J,43}^{(\gamma)}(1,2),
\label{eq:HJapp}
\end{equation}
and neglect ${\cal H}_{J,25}^{(\gamma)}(1,2)$ terms altogether.
This approximation is used in Sec. \ref{sec:model}.

\section{Orbital operators in the L-basis}

The starting point to express the orbital operators appearing in the
spin-orbital superexchange model (\ref{fullH}) is the relation between 
quenched
$\left|a\right\rangle_i$, $\left|b\right\rangle_i$, and
$\left|c\right\rangle_i$ 
orbitals at site $i$ and the eigenvectors $\left|1\right\rangle_i$,
$\left|0\right\rangle_i$, and $\left|-1\right\rangle_i$ of the 
angular momentum operator $L^z_i$. These are known to be
\begin{eqnarray}
\left|a\right\rangle_i& = & \frac{1}{\sqrt{2}}
\left(\left|1\right\rangle_i+\left|-1\right\rangle_i\right), \nonumber \\
\left|b\right\rangle_i  & = & \frac{-i}{\sqrt{2}}
\left(\left|1\right\rangle_i-\left|-1\right\rangle_i\right), \nonumber \\
\left|c\right\rangle_i  & = & \left|0\right\rangle_i .
\end{eqnarray}
From this we can immediately get the occupation number operators for
the doublon,
\begin{eqnarray}
D_i^{(a)} & =a^{\dagger}_ia_i^{}=\left|a\right\rangle_i\left\langle a\right|_i=
& 1-\left(L^{x}_i\right)^{2},\nonumber \\
D_i^{(b)} & =b^{\dagger}_ib_i^{}=\left|b\right\rangle_i\left\langle b\right|_i=
& 1-\left(L^{y}_i\right)^{2},\nonumber \\
D_i^{(c)} & =c^{\dagger}_ic_i^{}=\left|c\right\rangle_i\left\langle c\right|_i=
& 1-\left(L^{z}_i\right)^{2},
\end{eqnarray}
and the related $\{n_i^{(\gamma)}\}$ operators,
\begin{eqnarray}
n_i^{(a)} & =b^{\dagger}_ib_i^{}+c^{\dagger}_ic_i^{}=
& \left(L^{x}_i\right)^{2},\nonumber \\
n_i^{(b)} & =c^{\dagger}_ic_i^{}+a^{\dagger}_ia_i^{}=
& \left(L^{y}_i\right)^{2},\nonumber \\
n_i^{(c)} & =a^{\dagger}_ia_i^{}+b^{\dagger}_ib_i^{}=
& \left(L^{z}_i\right)^{2}.
\end{eqnarray}
The doublon hopping operators have a slightly different structure that
reflects their noncommutivity, i.e.,
\begin{eqnarray}
a^{\dagger}_ib_i^{}& = &
\left|a\right\rangle_i\left\langle b\right|_i=iL^{y}_iL^{x}_i, \nonumber \\
b^{\dagger}_ic_i^{}& = &
\left|b\right\rangle_i\left\langle c\right|_i=iL^{z}_iL^{y}_i, \nonumber \\
c^{\dagger}_ia_i^{}& = &
\left|c\right\rangle_i\left\langle a\right|_i=iL^{x}_iL^{z}_i.
\end{eqnarray}
These relations are sufficient to write the superexchange Hamiltonian
for the host-host and impurity-host bonds in the 
$\left\{ L^x_i,L^y_i,L^z_i\right\}$ operator basis for the orbital part. 
However, in practice it is more convenient to work with real operators 
$\left\{L^+_i,L^-_i,L^z_i\right\}$ rather than with the original ones, 
$\left\{ L^x_i,L^y_i,L^z_i\right\}$. Thus we write the final relations 
which we used for the numerical calculations in terms of these operators,
\begin{eqnarray}
D_i^{(a)} & = &-\frac{1}{4}\left[\left(L^+_i\right)^{2}
+\left(L^-_i\right)^{2}\right]+\frac12\left(L^z_i\right)^2,\nonumber \\
D_i^{(b)} & = &\hskip .2cm \frac{1}{4}\left[\left(L^+_i\right)^{2}
+\left(L^-_i\right)^{2}\right]+\frac12\left(L^z_i\right)^2,\nonumber \\
D_i^{(c)} & = & 1-\left(L^z_i\right)^{2},
\end{eqnarray}
for the doublon occupation numbers and going directly to the orbital
$\vec{\tau}_i$ operators we find that,
\begin{eqnarray}
\tau^{+(a)}_i& = & \frac12\left(L^-_i-L^+_i\right)L^{z}_i,  \nonumber \\
\tau^{+(b)}_i& = &\frac{-i}{2}\,L^z_i\left(L^+_i+L^{-}_i\right), \nonumber \\
\tau^{+(c)}_i& = & \frac{i}{4}\left[\left(L^{+}_i\right)^{2}
-\left(L^{-}_i\right)^{2}\right]-\frac{i}{2}L^{z}_i,
\end{eqnarray}
for the off-diagonal part and
\begin{eqnarray}
\tau^{z(a)}_i & = & \frac{1}{8}\left[\left(L^{+}_i\right)^{2}
+\left(L^{-}_i\right)^2\right]+\frac{3}{4}\left(L^{z}_i\right)^2
-\frac{1}{2},\nonumber \\
\tau^{z(b)}_i & = & \frac{1}{8}\left[\left(L^{+}_i\right)^{2}
+\left(L^{-}_i\right)^2\right]-\frac{3}{4}\left(L^{z}_i\right)^2
+\frac{1}{2},\nonumber \\
\tau^{z(c)}_i & = & -\frac{1}{4}\left[\left(L^{+}_i\right)^{2}
+\left(L^{-}_i\right)^{2}\right],
\end{eqnarray}
for the diagonal one. Note that the complex phase in $\tau^{+(b)}_i$
and $\tau^{+(c)}_i$ is irrelevant and can be omitted here as
$\tau^{+(\gamma)}_i$ is always accompanied by $\tau^{-(\gamma)}_j$ on a
neighboring site. This is a consequence of the cubic symmetry in the
orbital part of the superexchange Hamiltonian and it can be altered by
a presence of a distortion, e.g., octahedral rotation. For completeness
we also give the backward relation between angular momentum components,
$\{L^{\alpha}_i\}$ with $\alpha=x,y,z$, and the orbital operators
$\{\tau^{\alpha(\gamma)}_i\}$; these are:
\begin{eqnarray}
L^{x}_i & = & 2\tau^{x(a)}_i, \nonumber \\
L^{y}_i & = & 2\tau^{x(b)}_i, \nonumber \\
L^{z}_i & = & 2\tau^{y(c)}_i.
\end{eqnarray}

\end{document}